\documentclass[aps,preprintnumbers,floatfix,amsmath,amssymb,11pt]{revtex4}

\usepackage{graphicx}

\usepackage{subfigure}

\usepackage{bm}

\begin{document}


\def\Qs{Q_{\rm s}}
\def\half{{\textstyle{\frac12}}}
\def\p{{\bm p}}
\def\q{{\bm q}}
\def\x{{\bm x}}
\def\v{{\bm v}}
\def\E{{\bm E}}
\def\B{{\bm B}}
\def\A{{\bm A}}
\def\a{{\bm a}}
\def\b{{\bm b}}
\def\c{{\bm c}}
\def\j{{\bm j}}
\def\n{{\bm n}}
\def\grad{{\bm\nabla}}
\def\da{d_{\rm A}}
\def\tr{\operatorname{tr}}
\def\Im{\operatorname{Im}}
\def\Re{\operatorname{Re}}
\def\log{\operatorname{log}}
\def\md{m_{\rm D}}
\def\mpl{m_{\rm pl}}
\def\pol{\varepsilon}
\def\bpol{{\bm\pol}}
\def\CP{CP^{N-1}}



\title
    {%
      Small Instantons in $CP^1$ and $CP^2$ Sigma Models
    }%

\author {Yaogang Lian and H. B. Thacker}

\affiliation
    {%
    Department of Physics,
    University of Virginia,
    P.O. Box 400714
    Charlottesville, VA 22901-4714
    }%

\date {\today}

\begin {abstract}%
    {%
The anomalous scaling behavior of the topological susceptibility $\chi_t$
in two-dimensional $CP^{N-1}$ sigma models for $N\leq 3$ is studied using the overlap
Dirac operator construction of the lattice topological charge density. The divergence of $\chi_t$ in
these models is traced to the presence of small instantons with a radius of order $a$ 
(= lattice spacing), which are directly observed on the lattice. 
The observation of these small instantons provides detailed confirmation of L\"{u}scher's argument that such 
short-distance excitations, with quantized topological charge, 
should be the dominant topological fluctuations in $CP^1$ and $CP^2$,
leading to a divergent topological susceptibility in the continuum limit. For the $\CP$ models with $N>3$
the topological susceptibility is observed to scale properly with the mass gap.
These larger $N$ models are not dominated by instantons, but rather by coherent, one-dimensional regions of topological charge
which can be interpreted as domain wall or Wilson line excitations and are analogous to D-brane or ``Wilson bag''
excitations in QCD. In Lorentz gauge, the small instantons
and Wilson line excitations can be described, respectively, in terms of poles and cuts 
of an analytic gauge potential.
    }%
\end {abstract}

\maketitle
\thispagestyle {empty}

\section {Introduction}
\label{sec:intro}
Recent lattice studies of topological charge structure in pure-glue QCD have provided 
evidence that the topological charge density, as constructed from the overlap Dirac operator, 
organizes into low-dimensional long-range structure \cite{Horvath:2003yj}, consisting of
3-dimensional coherent sheets in 4-dimensional spacetime. A 
further study demonstrated that this structure is inherently global and that topological 
charge fluctuations in QCD should be understood not in terms of individual localized lumps 
but rather as extended brane-like objects \cite{Horvath:2005rv,Horvath-contact}. 
These lattice results challenge the traditional instanton gas or liquid picture 
\cite{Callan:1977gz}, in which the QCD vacuum is described as an ensemble of 
localized (anti-)self-dual lumps of quantized topological charge. On the other hand,
the results are quite consistent with the view of QCD theta-dependence provided by string/gauge
duality \cite{Witten98} \cite{Aharony:1999ti}, in which topological charge excitations come in the form of membranes or
domain walls \cite{ShifmanGabadadze,Shifman}, which are the gauge theory dual of wrapped
6-branes in type IIA string theory. Although the nature of 
topological charge excitations in QCD is far from a settled issue, lattice studies may 
shed light on the dynamics of these coherent fluctuations and their relation to 
confinement and spontaneous chiral symmetry breaking. Analogous long-range 
sign-coherent 1-dimensional regions of topological charge density have also been found recently in 
a Monte Carlo study of 2D $\CP$ sigma models \cite{Ahmad}. Two-dimensional $\CP$ 
models and four-dimensional QCD have many properties in common. In particular, both are 
classically scale invariant and have classical instanton solutions of arbitrary radius. 
Lattice evidence for the existence of long-range coherent structures of codimension one
in both QCD and $\CP$ strongly supports the understanding of QCD topological charge excitations
in terms of membranes. In 
both of these models it has been argued \cite{Ahmad,lat05} that the observed membranes are associated with boundaries 
between ``k-vacua'', characterized by an effective local value of $\theta$ which jumps 
by $\pm2\pi$ across the boundary. The presence of such domain walls in the QCD vacuum was suggested
much earlier on the basis of large $N$ chiral symmetry arguments  \cite{Witten79}. 
The same kind of 3D membrane structure in QCD was also identified by L\"{u}scher \cite{Luscher78}
in the form of ``Wilson bags'' which are 3-dimensional surface integrals over the 3-index Chern-Simons
tensor and are analogous to Wilson lines in the $\CP$ models.

In Ref.~\cite{Ahmad} the scaling behavior of the overlap-based topological susceptibility
was studied for $CP^1$, $CP^3$, and $CP^9$. For $CP^3$ and $CP^9$, $\chi_t$ was found 
to scale properly with the mass gap. However, for $CP^1$,
$\chi_t$ exhibits anomalous scaling behavior, giving an apparently divergent 
topological susceptibility in the continuum limit. This reproduced a result first 
obtained by Berg and L\"{u}scher in a Monte Carlo
study of the lattice $O(3)$ sigma model \cite{BergLuscher}. The topological susceptibility 
$\chi_t = \langle Q^2 \rangle/V$ ($Q$= total topological charge, $V$= spacetime volume) in two-dimensional 
$\CP$ sigma models would normally be expected to be a renormalization group 
invariant parameter, which 
should scale like the squared mass gap $\mu^2$ ($\mu=$ mass of nonsinglet meson) for large $\beta$ ($\beta=1/g^2$, $g$ is the bare 
coupling constant). However, the Monte Carlo results reported in Ref.~\cite{Ahmad} 
show that $\chi_t$ does not even 
approximately scale in the case of $CP^1$, although it does scale properly for $CP^3$ and $CP^9$. 
L\"{u}scher \cite{Luscher82} showed that the anomalous scaling behavior of $\chi_t$ 
for $CP^1$ could be explained by the contribution of small, point-like instantons with a radius of order $a$ 
(= lattice spacing). Although the action of a classical continuum instanton is 
independent of its radius, on the lattice this is not the case, and for $CP^1$ and $CP^2$,
the small instantons have such a low action that they overwhelm 
the contribution of more slowly varying fields which might have been expected to 
dominate in the continuum limit. 
This mechanism is at work for $CP^1$ and $CP^2$, but not for $CP^3$ and higher $\CP$ 
models, because the action of the small instantons is proportional to $N$ and 
becomes too large to be favored for $N \ge 4$. (As we will show, $CP^3$ is a marginal case where instantons are observed but do
not dominate as in $CP^1$ and $CP^2$.) The abrupt change in the nature of the
dominant topological charge fluctuations as $N$ is increased above 4 provides a detailed 
Monte Carlo example of the phenomenon first discussed by Witten (for both QCD and $\CP$)\cite{Witten79}, in which instanton
effects are exponentially suppressed at large $N$, and topological charge is dominated by
quantum effects. This can be described, at least figuratively, by saying
that instantons ``melt'' due to quantum fluctuations inherent in the large $N_c$ confining vacuum.
For the case of the $\CP$ models, the instanton ``melting point'' is $N\approx 4$.  

The definition of the lattice topological charge density used in the original
study by Berg and L\"{u}scher was an ultralocal discretization of the topological 
charge operator in the $O(3)$ sigma model formulation.\cite{BergLuscher} For
our formulation of the
lattice $\CP$ models, the simplest ultralocal definition of topological charge is 
the log of the plaquette constructed from the $U(1)$ gauge links,
\begin{equation}
\label{eqn:q_plaq}
q_p(x) = \frac{1}{2\pi} \Im \{\log [U_{\mu}(x)U_{\nu}(x+\hat{\mu})U_{\mu}^{\dagger}(x+\hat{\mu}+\hat{\nu})U_{\nu}^{\dagger}(x+\hat{\nu}) ]\}
\end{equation}
This operator is ultralocal in the gauge fields. It provides an adequate lattice 
definition of global topological charge and gives results for topological susceptibility
which are similar but not identical to those obtained from the overlap formalism. 
But as a definition of local topological 
charge density the plaquette-based $q_p(x)$ has severe shortcomings. It suffers 
from a large amount of short-range noise associated with local gauge field fluctuations, 
which obscures the long-range coherent topological charge structures found with the
overlap-based topological charge density. The overlap 
definition of the lattice topological charge density is \cite{Hasenfratz:1998ri}
\begin{equation}
\label{eqn:q_overlap}
q(x) = \frac{1}{2} \tr [\gamma_5 D(x,x)]
\end{equation}
where $D$ is a lattice Dirac operator (in our case the overlap operator \cite{Neuberger})
which satisfies the Ginsparg-Wilson relation
\begin{equation}
\label{eqn:gw}
\{D, \gamma_5\} = a D \gamma_5 D
\end{equation}
This construction of the topological charge density on the lattice provides a more incisive 
probe into the topological structure of gauge field configurations than the plaquette-based
operator (\ref{eqn:q_plaq}).  
The exact lattice chiral symmetry of the overlap Dirac operator leads to an implicit ``chiral
smoothing'' of the topological charge which takes place over distances of order $a$.
This reveals the coherent structure of the topological fluctuations without
any subjective cooling or smoothing procedure.

In this paper, we investigate the origin of the anomalous 
scaling behavior of topological susceptibility 
in two-dimensional $CP^1$ and $CP^2$ sigma models, using the overlap construction of 
the lattice topological charge density. We directly observe small instantons with 
radii of order $a$ (roughly one to two lattice spacings) in Monte Carlo configurations for
both $CP^1$ and $CP^2$. In the continuum (large $\beta$) limit these small instantons
dominate the topological susceptibility. 
In contrast to these point-like small instanton excitations,
the topological charge in $\CP$ models for $N\ge 4$ has been found to be 
dominated by one-dimensional line-like ``domain wall'' structures \cite{Ahmad}.
After establishing the prevalence of small instantons in $CP^1$ and $CP^2$,
we measure the action of an individual small instanton and show that
the measured topological susceptibility is in good agreement with 
a dilute instanton gas calculation.
Finally we discuss a description of topological fluctuations in Lorentz gauge 
which incorporates both small instantons 
and domain wall structures into a coherent picture in terms of poles and cuts 
in an analytic gauge potential.

\section {Anomalous Scaling Behavior of Topological Susceptibility}
\label{sec:anomalous}
\subsection { Lattice $\CP$ Sigma Models}
\label{anomalous:cpn}
The Lagrangian for $\CP$ sigma models in the continuum is
\begin{equation}
\label{eqn:L_cpn1}
L = \partial_{\mu}z_i^*\partial_{\mu}z_i + (z_i^*\partial_{\mu}z_i)(z_j^*\partial_{\mu}z_j)
\end{equation}
where $z^i$ are N complex fields, $i=1,\dots,N$, satisfying a constraint $z_{i}^{*}z_{i}=1$.
This Lagrangian is invariant under a local U(1) gauge transformation: $z_{i}(x)\to e^{ia(x)}z_{i}(x)$, 
for arbitrary space-time dependent $a(x)$. We can introduce a dummy gauge field $A_{\mu}$ and 
rewrite the Lagrangian as
\begin{equation}
\label{eqn:L_cpn2}
L = (\partial_{\mu} - iA_{\mu})z_{i}^{*}(\partial_{\mu} + iA_{\mu})z_i
\end{equation}
where 
\begin{equation}
\label{A_mu_cpn0}
A_{\mu} = \frac{1}{2}i (\mathbf{z}^{\dagger} \partial_{\mu} \mathbf{z} - (\partial_{\mu} \mathbf{z})^{\dagger} \mathbf{z})
\end{equation}

To put $\CP$ models on the lattice, we introduce $U(1)$ link fields $U(x, x+\hat{\mu}) = e^{iA_{\mu}(x)}$.
$\CP$ fields are defined on the sites as $\mathbf{z}(x)$. The lattice action 
consists of gauge-invariant nearest-neighbor hopping terms,
\begin{equation}
\label{eqn:cpn_action}
S = \beta N \sum_{x, \hat{\mu}} \mathbf{z}(x)^\dagger U(x, x+\hat{\mu}) \mathbf{z}(x+\hat{\mu}) + c.c.
\end{equation}
This lattice action is used in our Monte Carlo simulation to generate an ensemble of field configurations. 
The $\mathbf{z}$ fields are updated by a Cabibbo-Marinari heat bath algorithm, while $U(1)$ link fields are updated by 
a multi-hit Metropolis algorithm.

\subsection {Overlap Dirac Operator}
\label{anomalous:overlap}
The overlap Dirac operator \cite{Neuberger} preserves exact chiral symmetry on the lattice via the Ginsparg-Wilson
relation. For the present work, we employ a standard construction starting 
from a Wilson-Dirac kernel $D_{W}$. The overlap Dirac operator can be written as
\begin{equation}
\label{eqn:overlap_from_Hw}
D = \frac{1}{a} (1+ \gamma_5 \epsilon(H_{W}(m)))
\end{equation}
where $H_{W}(m) = \gamma_5 D_{W}(-m)$ and 
\begin{equation}
\label{eqn:epsilon_Hw}
\epsilon(H_{W}(m)) = \frac{H_{W}(m)}{\sqrt{H_{W}^{\dagger}(m) H_{W}(m)}}
\end{equation}
It can be easily verified that this construction of the overlap Dirac operator satisfies the Ginsparg-Wilson 
relation, Eq. (\ref{eqn:gw}).

Using the overlap Dirac operator, we can define the lattice topological charge density \cite{Hasenfratz:1998ri} by 
Eq. (\ref{eqn:q_overlap}),
where the trace is summed over spin indices in $\CP$ models and over spin and color indices in QCD. 
The overlap definition of topological charge is not ultralocal in the gauge links. This has the effect of performing some
smoothing of the gauge field over a range of order $a$. For smooth $U(1)$ gauge fields in two dimensions, 
we find that the overlap
topological charge at a site is reasonably well-approximated by the average of the four 
plaquette-based charges surrounding the site. For a rapidly varying field, and particularly for the small
instantons that we observe in $CP^1$ and $CP^2$, the overlap-based charge provides a much clearer view
of coherent excitations. 
The effect of the smoothing obtained by the overlap construction can also be seen in the structure of the 2-point topological
charge correlator. Because it is ultralocal, reflection positivity arguments imply that the nearest-neighbor 
correlator for the plaquette-based charge must be negative. On the other hand, the overlap-based nearest-neighbor
correlator is found to be positive. This allows the overlap-based operator to detect sign-coherent structure even
if that structure is in the form of small $O(a)$-radius instantons or membranes of $O(a)$ thickness.

\subsection {Topological Susceptibility}
\label{anomalous:topological_susceptibility}
Using the overlap definition of the lattice topological charge density, we have studied the scaling 
behavior of topological susceptibility $\chi_t$ in $CP^1$, $CP^2$, $CP^3$, $CP^5$, and $CP^9$ sigma models. 
Since $\chi_t$ is generally expected to be a renormalization group invariant parameter in two-dimensional $\CP$ 
sigma models, it is expected to scale like a $(mass)^2$ in the continuum limit. There is a natural mass 
scale in these models which is the mass gap of the $z_{i}^{*}z_{j}, (i\neq j)$ meson correlator in the nonsinglet channel. The
zero-momentum meson correlator falls off exponentially at large Euclidean time, 
\begin{equation}
\label{eqn:meson_correlator}
\int dx_2 \langle z_{i}^{*}(x)z_{j}(x)z_{j}^{*}(0)z_{i}(0) \rangle \sim const. \times e^{-\mu x_1}
\end{equation}
where $\mu$ is the mass gap.

We have measured the mass gap $\mu$ in lattice units for various values of $N$ and $\beta$ as shown 
in Table~\ref{tab:meson_mass}.
\begin{table}[ht]
\begin{center}
\caption{\label{tab:meson_mass} Mass gap of the meson correlator in lattice units}
\begin{tabular}{||c|c||c|c||c|c||c|c||c|c||}
\hline
\hline
$\beta$ & $CP^1$ & $\beta$ & $CP^2$ & $\beta$ & $CP^3$ & $\beta$ & $CP^5$ & $\beta$ & $CP^9$\\
\hline
\hline
1.0 & .438(5) & 0.9 & .436(5) & 0.8 & .554(2) & 0.7 & .652(6) & 0.7 & .406(2)\\
\hline
1.1 & .286(5) & 1.0 & .275(2) & 0.9 & .327(3) & 0.8 & .360(3) & 0.8 & .212(2)\\
\hline
1.2 & .179(3) & 1.1 & .155(1) & 1.0 & .180(1) & 0.9 & .186(3) & 0.9 & .0895(6)\\
\hline
1.3 & .111(1) & 1.2 & .084(2) & 1.1 & .088(1) & 1.0 & .085(1) & 1.0 & .0579(2)\\
\hline
1.4 & .070(1) & 1.3 & .043(1) & 1.2 & .0531(3) & 1.1 & .0496(9) & 1.1 & .0475(4) \\
\hline
1.5 & .036(1) & 1.4 & .0256(8) & 1.3 & .0264(7) & 1.2 & .0308(4) & 1.2 & .0287(4)\\
\hline
1.6 & .0238(8) & 1.5 & .0207(5) & 1.4 & .0217(5) & 1.3 & .0249(2) & 1.3 & .0255(3)\\
\hline
\hline
\end{tabular}
\end{center}
\end{table}
We worked on lattice sizes up to $100\times100$. For each $N$, 
$\beta$ has been chosen to cover the region where correlation 
length $\xi=1/\mu$ varies from roughly 3 to 50. For each set of $N$ and $\beta$, the meson correlator was averaged 
over $4000$ field configurations. The correlator fits were carried out by a standard covariant $\chi^2$
minimization. 

\begin{figure}
  \centering
  \includegraphics[width=0.65\textwidth, angle=-90]{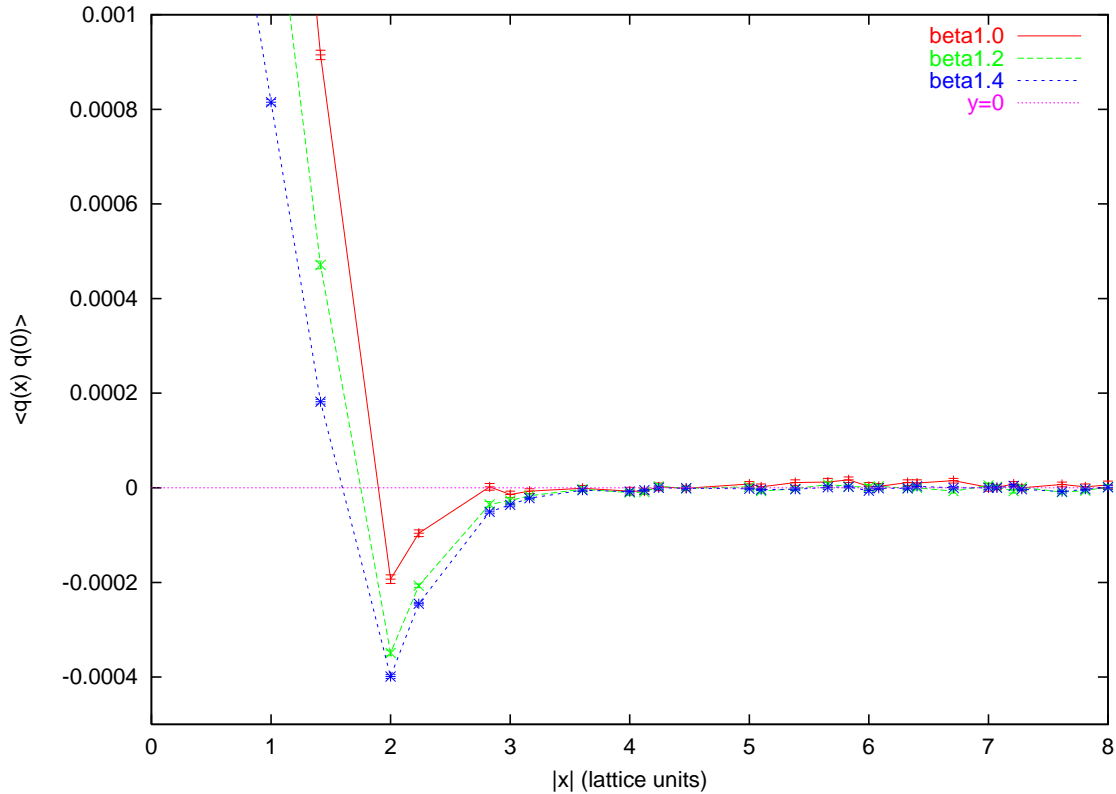}
  \caption{ \label{fig:TCcorrelators} The two-point topological charge density correlator for $CP^1$ with several values of $\beta$.}
\end{figure} 

\begin{figure}
  \centering
  \includegraphics[width=0.65\textwidth, angle=-90]{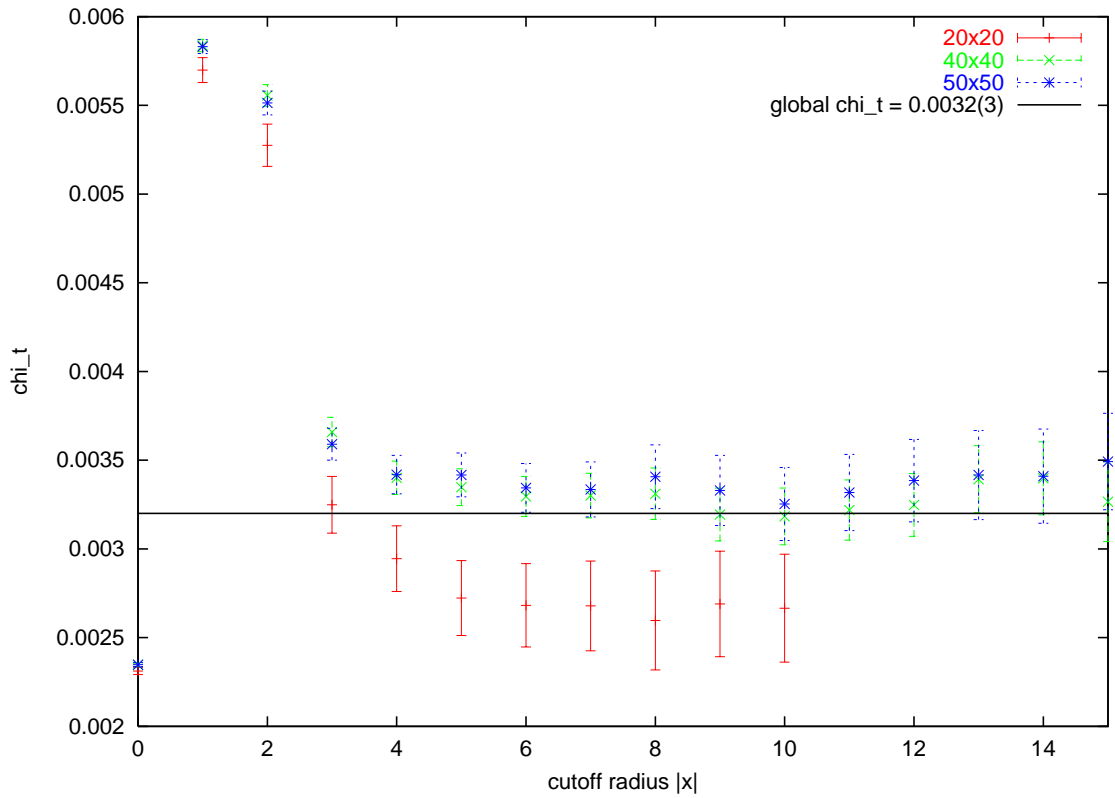}
  \caption{ \label{fig:cp1_b10_integ} $\chi_t$ vs. cutoff radius}
\end{figure}

The topological susceptibility $\chi_t$ can be determined from the large-$V$ limit of
\begin{equation}
\label{eqn:chi_t_vol}
\chi^{(V)}_{t} = \langle Q^2 \rangle/V
\end{equation}
where $Q$ is the total topological charge on the lattice and $V$ is the space-time volume. 
An equivalent way to measure $\chi_t$ is to integrate the 
two-point topological charge density correlator,
\begin{equation}
\label{eqn:chi_t_cor}
\chi_{t}= \sum_x \langle q(x)q(0) \rangle
\end{equation}
where $q(x)$ is calculated from the overlap definition (Eq.~\ref{eqn:q_overlap}). 
If the correlator calculation is carried out at finite volume, and the 
integral is taken over the entire volume of the lattice, this gives results identical
to those obtained by measuring the global charge $Q$ of each configuration and using (\ref{eqn:chi_t_vol}).
However, we have found that at large values of $\beta$, where the correlation length $\mu^{-1}$ is long and finite
volume effects become important, a method based on a truncated integration of the correlator is much less affected
by finite volume corrections than the calculation based on measurement of the global charge $Q$.
The infinite volume topological charge correlator is very short range, consisting of a positive
core at $x<2$ and a short-range negative tail which falls off rapidly, even when the nonsinglet meson correlation
length $\mu^{-1}$ is quite long. Fig. \ref{fig:TCcorrelators} shows the point-to-point correlator 
$\langle q(x)q(0)\rangle$ for $CP^1$ for several different values of $\beta$. Over this same range, 
$\mu^{-1}$ varies from 2.3 to 14.4. For larger values of $\beta$,
where the correlation length becomes comparable to the box size, the global charge $Q$ freezes to zero.
The integrated correlator also goes to zero, but, at least for correlation lengths up to about half the box size,
the shape and magnitude of the short-range topological charge correlator remains relatively unchanged from its
infinite volume value. The fact that the integrated finite volume correlator is smaller comes about through a cancellation between
the short range part of the integral (which still gives a reasonable approximation to the infinite volume 
susceptibility) and a negative background contribution which does not fall off as rapidly with distance and therefore gets
most of it's contribution from larger $|x|$. By comparing the correlator on several box sizes, we find that
this longer range negative background is a finite volume effect. As a result, on smaller boxes
the integrated short-range correlator gives a 
better estimate of the infinite volume $\chi_t$ than one could obtain from the fluctuations of the global charge.
After experimenting with different prescriptions for cutting off the short-range correlator integration, we have
chosen to calculate the correlator as a function of $|x|$ and integrate out to a cutoff value of $|x|\leq x_c$ 
for which the measured correlator is within one standard deviation of zero. This was always in the range $3<x_c<5$.
Results were essentially unchanged if we used a constant cutoff of $x_c=4$.
To test the method, we calculated $\chi_t$ as a function of the cutoff radius for $CP^3$ at $\beta=1.0$ on 
various lattice sizes, as shown in Fig.~\ref{fig:cp1_b10_integ}. $\chi_t$ measured on the $40\times40$ lattice 
matches nicely with that measured on the $50\times50$ lattice at each cutoff radius. This shows that the 
$50\times50$ lattice is large enough to give a correct $\chi_t$ that is close to the infinite volume susceptibility. 
By using (\ref{eqn:chi_t_vol}) on the $50\times50$ lattice configurations, we obtained $\chi_t = 0.0032(3)$. 
From the graph, we observe that $\chi_t$ measured on the $40\times40$ or $50\times50$ lattice quickly converges to the 
infinite volume result when $x_c\approx 4$. However, on the $20\times20$ lattice, $\chi_t$ converges to a 
value lower than the infinite volume result. On the other hand, if we apply the cutoff at $x_c\approx 4$, we can 
still extract approximately the right value of $\chi_t$ on the $20\times20$ lattice. Most of the finite volume effect that causes 
the globally determined $\chi_t$ to be too small on the $20\times20$ lattice comes from distances $>4$.
On the larger $40\times 40$ and $50\times 50$ lattices, the value obtained for $\chi_t$ is essentially independent
of the cutoff for $x_c>4$, indicating that the longer range negative contribution from $|x|>4$ on the smaller box
is a finite volume effect. This shows that the truncated integration 
method is less affected by finite volume than a calculation based on measurement of the global charge $Q$.
Using the truncated integration method, we have measured $\chi_t$ in lattice units for various values of $N$ and $\beta$, 
as shown in Table~\ref{tab:chi_t}.

\begin{table}[ht]
\begin{center}
\caption{\label{tab:chi_t} Topological susceptibility $\chi_t$ in lattice units}
\begin{tabular}{||c|c||c|c||c|c||c|c||c|c||}
\hline
\hline
$\beta$ & $CP^1$ & $\beta$ & $CP^2$ & $\beta$ & $CP^3$ & $\beta$ & $CP^5$ & $\beta$ & $CP^9$\\
\hline
\hline
1.0 & .0146(2) & 0.9 & .0137(2) & 0.8 & .0147(2) & 0.7 & .0154(3) & 0.7 & .0051(1)\\
\hline
1.1 & .0106(2) & 1.0 & .0081(2) & 0.9 & .0079(2) & 0.8 & .0069(1) & 0.8 & .00105(5)\\
\hline
1.2 & .0070(2) & 1.1 & .0040(1) & 1.0 & .0034(1) & 0.9 & .0021(1) & 0.9 & .00021(2)\\
\hline
1.3 & .0040(1) & 1.2 & .00157(7) & 1.05 & .00194(8) & 1.0 & .00027(4) & 1.0 & .00009(2)\\
\hline
1.4 & .0022(1) & 1.3 & .00056(5) & 1.1 & .00100(6) &  1.1 & N/A & 1.1 & N/A \\
\hline
1.5 & .00104(5) & 1.4 & .00035(3) & 1.15 & .00057(5) & 1.2 & N/A & 1.2 & N/A \\
\hline
1.6 & .00053(5) & 1.5 & N/A & 1.2 & .00026(4) & 1.3 & N/A & 1.3 & N/A \\
\hline
\hline
\end{tabular}
\end{center}
\end{table}

\begin{figure}
  \centering
  \includegraphics[width=0.65\textwidth, angle=-90]{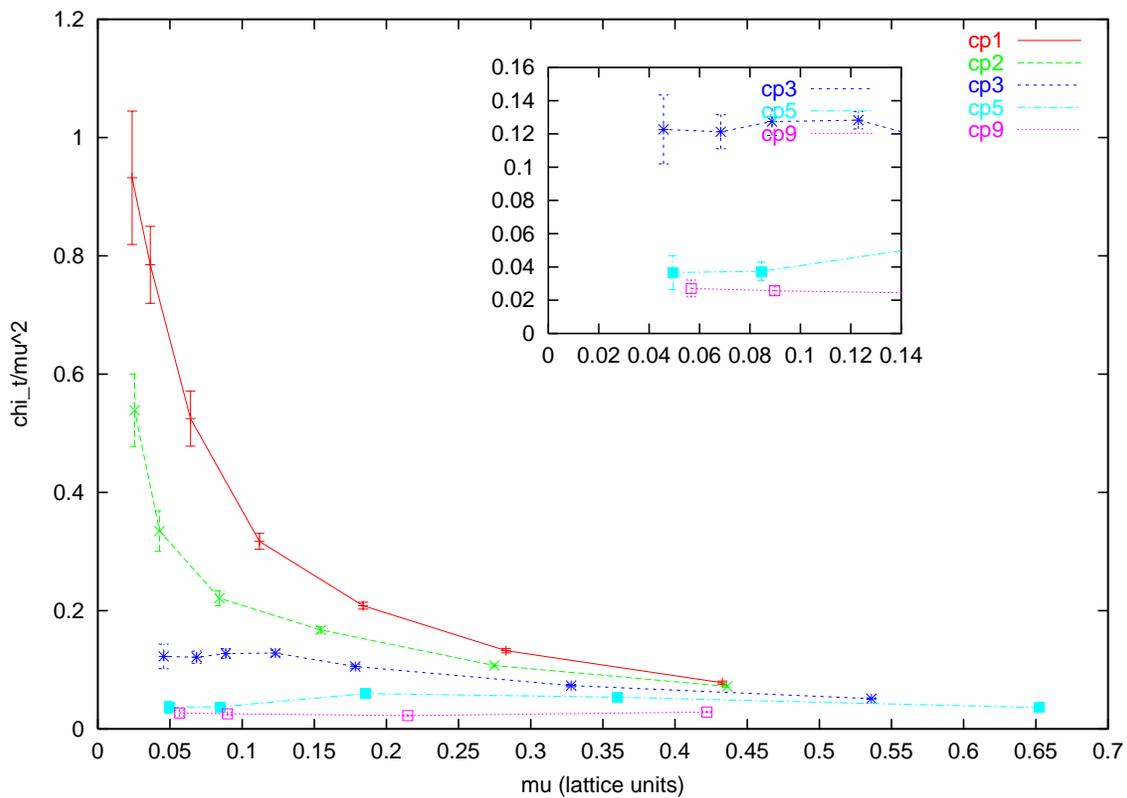}
  \caption{\label{fig:chi_t_scaling} Scaling behavior of $\chi_{t}/\mu^2$ in $CP^1$, $CP^2$, $CP^3$, $CP^5$ and $CP^9$}
\end{figure}

To show the scaling behavior of $\chi_t$ with respect to $(mass)^2$, we plot $\chi_t/\mu^2$ 
vs. $\mu$ in Fig.~\ref{fig:chi_t_scaling}. A constant value on the graph indicates proper scaling.
From the graph we observe that $\chi_t$ appears to scale properly for $CP^3$, $CP^5$, and $CP^9$ in the continuum limit, 
but it does not even approximately scale for $CP^1$ and $CP^2$. Our results confirm that the overlap-based $\chi_t$
exhibits the same anomalous scaling behavior observed in the $O(3)$ sigma model
formulation of $CP^1$ \cite{BergLuscher}. From Fig.~\ref{fig:chi_t_scaling} 
we see that this anomalous scaling behavior occurs in $CP^1$ and $CP^2$, but not in $CP^3$, $CP^5$, or $CP^9$. This is exactly
the behavior predicted by L\"{u}scher \cite{Luscher82}, who showed that, for $N\leq 3$,
the topological susceptibility should be dominated by small instantons with a radius of order $a$.
In the next section, we will discuss the theoretical explanation of the anomalous 
scaling behavior of $\chi_t$, present the direct observation of small instantons in Monte Carlo configurations, 
and discuss some properties of these objects.

\section{Small Instantons}
\label{sec:small_instantons}
The divergent behavior of $\chi_t$ in $CP^1$ and $CP^2$ can be attributed to instantons 
with a radius of approximately one to two lattice spacings. These small instantons are short-range gauge field 
fluctuations with quantized 
topological charge. In the space of lattice gauge field configurations they reside near the boundary between the $Q=1$ 
and $Q=0$ sectors, in the sense that a small adjustment of the gauge links will cause the instanton to ``fall
through the lattice'' and turn into a $Q=0$ configuration. The dominance of these small instanton configurations
violates the usual assumptions about the continuum limit.
In a naive view of the continuum limit, one would assume that configurations are dominated by fields for which the values
of gauge-invariant local densities (e.g. action density or topological charge density) are smooth
over distances of order $a$.
Although the probability for the small instantons to occur is exponentially small, the probability for more smoothly
varying fields to have nonzero total topological charge is also exponentially small. Which type of fluctuation 
dominates the topological susceptibility depends on the precise value of the instanton action.

If we only take into account the contribution from the slowly varying fields, the topological susceptibility should 
scale like a $(mass)^2$ in the continuum limit, i.e.
\begin{equation}
\label{eqn:chi_t_slow}
\chi_{t}^{s.v.} \propto \beta^{4/N} e^{-4\pi\beta}, \quad \quad \left(\beta \to \infty \right)
\end{equation}
However, if we take into account the contribution from small instantons, a dilute gas calculation yields
\begin{equation}
\label{eqn:chi_t_dg}
\chi_{t}^{d.g.} \propto \beta^{-1} e^{-\beta\epsilon}, \quad \quad \left(\beta \to \infty \right)
\end{equation}
where $\beta\epsilon$ is the minimal action of gauge field configurations in the $Q=1$ sector.

In Ref.~\cite{Luscher82} the minimum-action field configuration with $Q=1$ was constructed and its action was
calculated. This gave the value of $\epsilon$ in the $O(3)$ model as $\epsilon = 6.69\dots$ 
For other $\CP$ models, it was argued that the minimal action of a small instanton should be proportional to $N$, 
according to the inherent factor of $N$ in the $\CP$ action (see Eq.~\ref{eqn:cpn_action}). Since $\epsilon < 4\pi$ 
for $CP^1$ and $CP^2$, the small instantons make the dominant contribution to the topological susceptibility. This 
explains why $\chi_t$ does not scale as $\mu^2$ in the continuum limit in $CP^1$ and $CP^2$. However, for $\CP$ 
models with $N>3$,  $\epsilon = \frac{N}{2} \cdot 6.69 > 4\pi$, so $\chi_t$ is dominated by less singular topological
fluctuations. On the other hand, even for $N>3$, the dominant excitations are not smooth distributions of topological
charge, but rather, singular domain-wall type excitations which give rise to a dominant contact term in the 2-point
correlator \cite{Ahmad}.  

\begin{figure}
\hspace{-0.5in}
  \centering
  \subfigure[$CP^1$, $\beta=2.5$, $Q=1.0$]{
    \label{fig:cp1_b25_Q1}
    \includegraphics[width=0.40\textwidth]{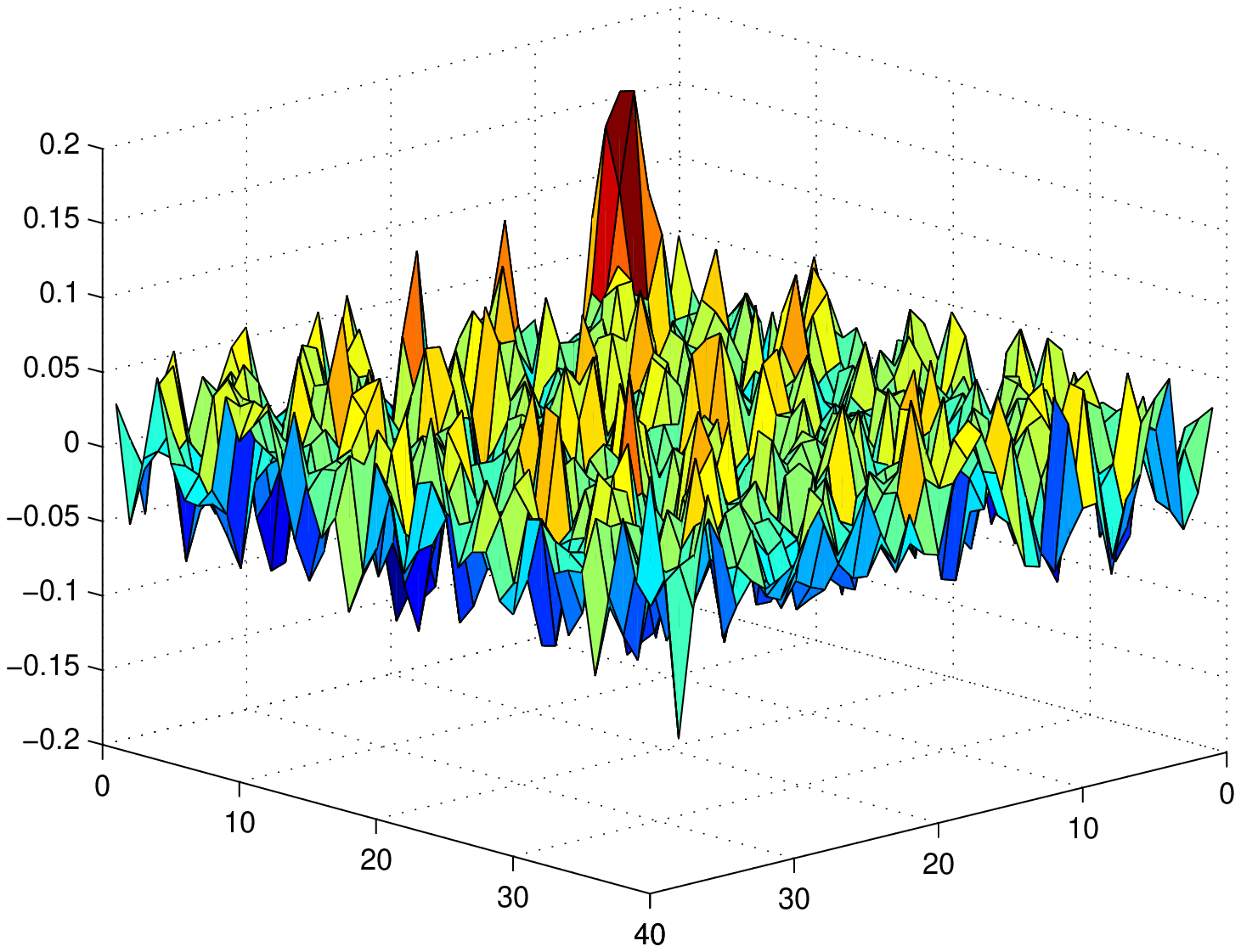}}
  \hspace{0in}
  \subfigure[$CP^1$, $\beta=2.5$, $Q=0$]{
    \label{fig:cp1_b25_Q0}
    \includegraphics[width=0.40\textwidth]{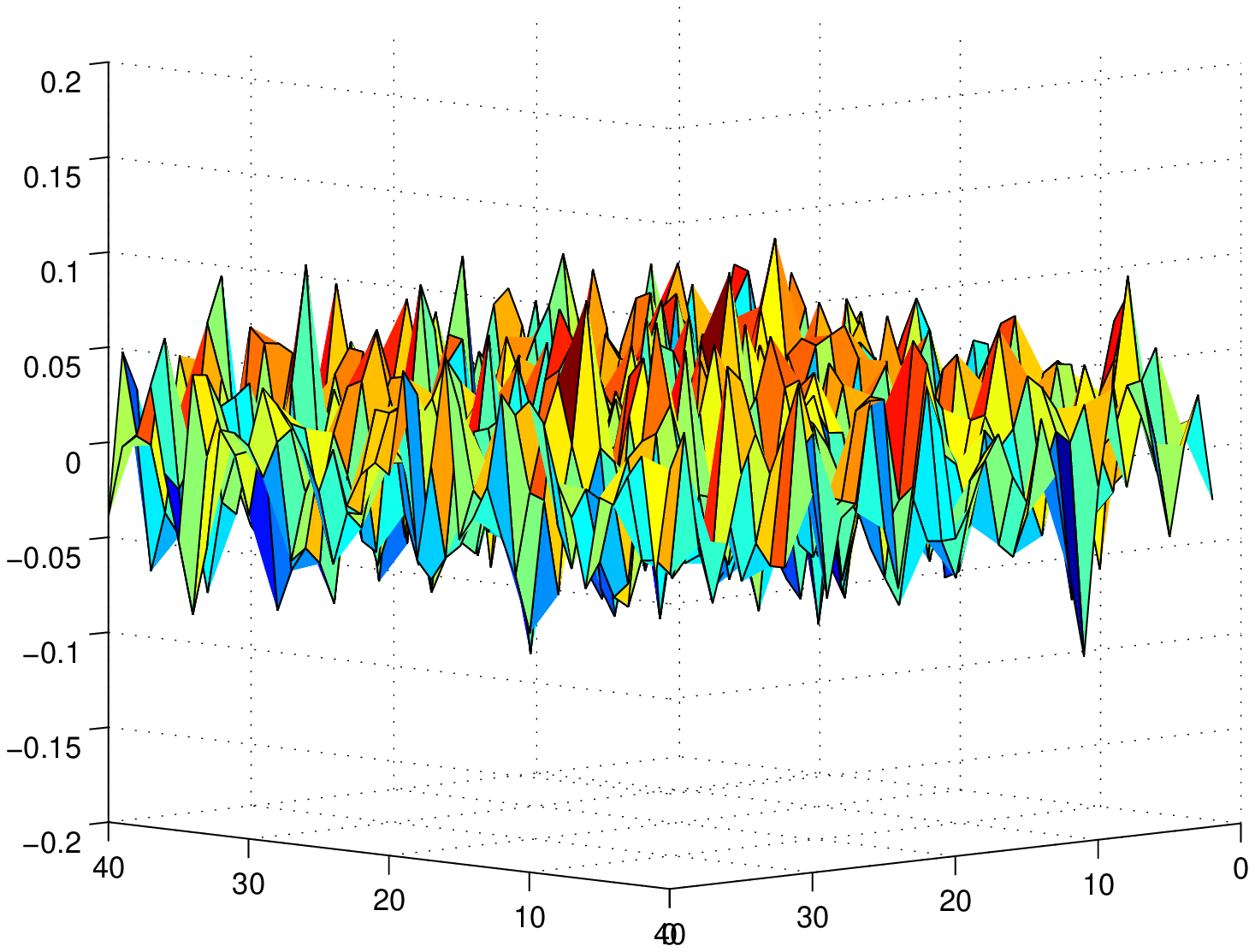}}\\
  \vspace{0in}
  \subfigure[$CP^1$, $\beta=1.6$, $Q=-2.0$]{
    \label{fig:cp1_b16}
    \includegraphics[width=0.40\textwidth]{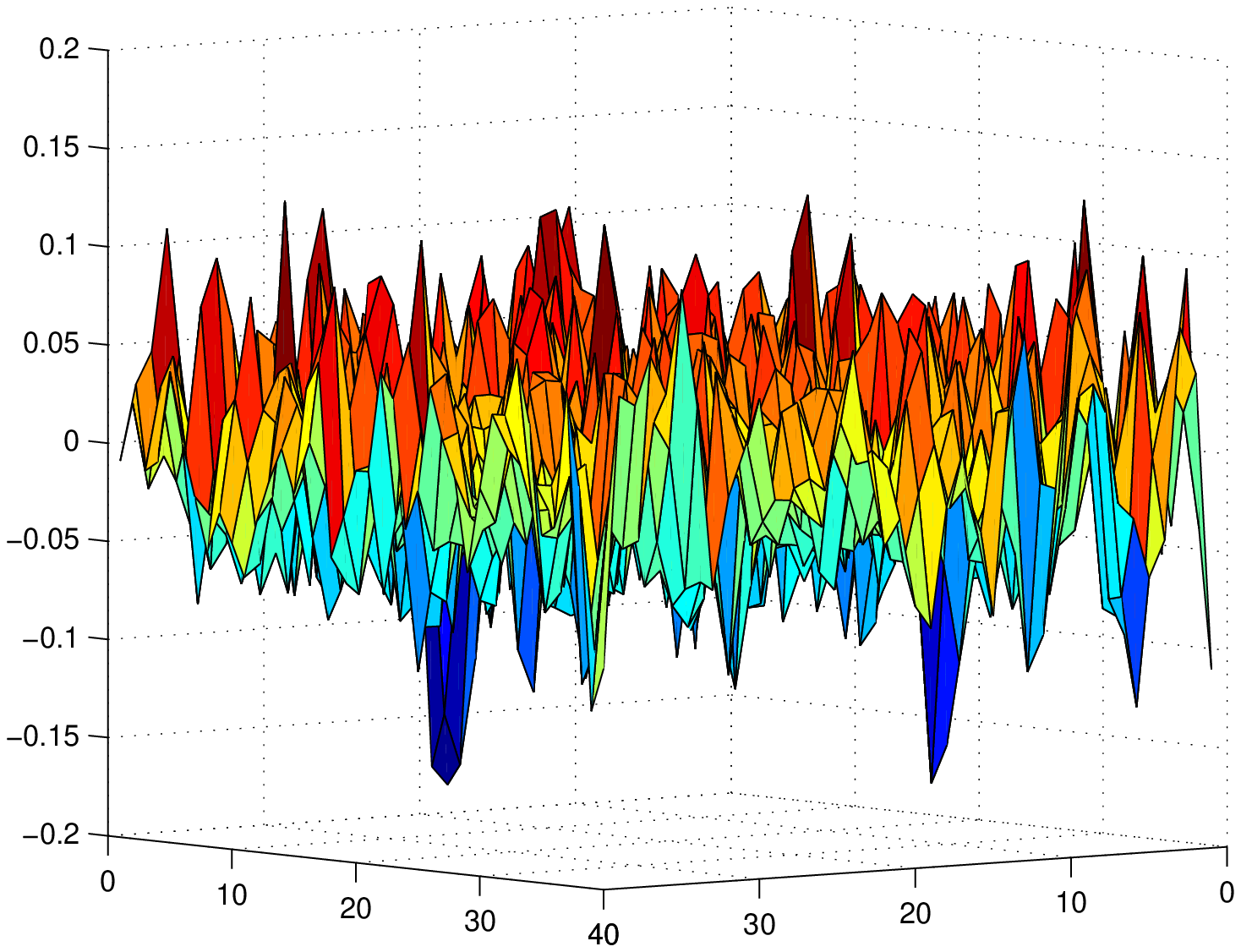}}
  \hspace{0in}
  \subfigure[$CP^2$, $\beta=1.8$, $Q=1.0$]{
    \label{fig:cp2_b18} 
    \includegraphics[width=0.40\textwidth]{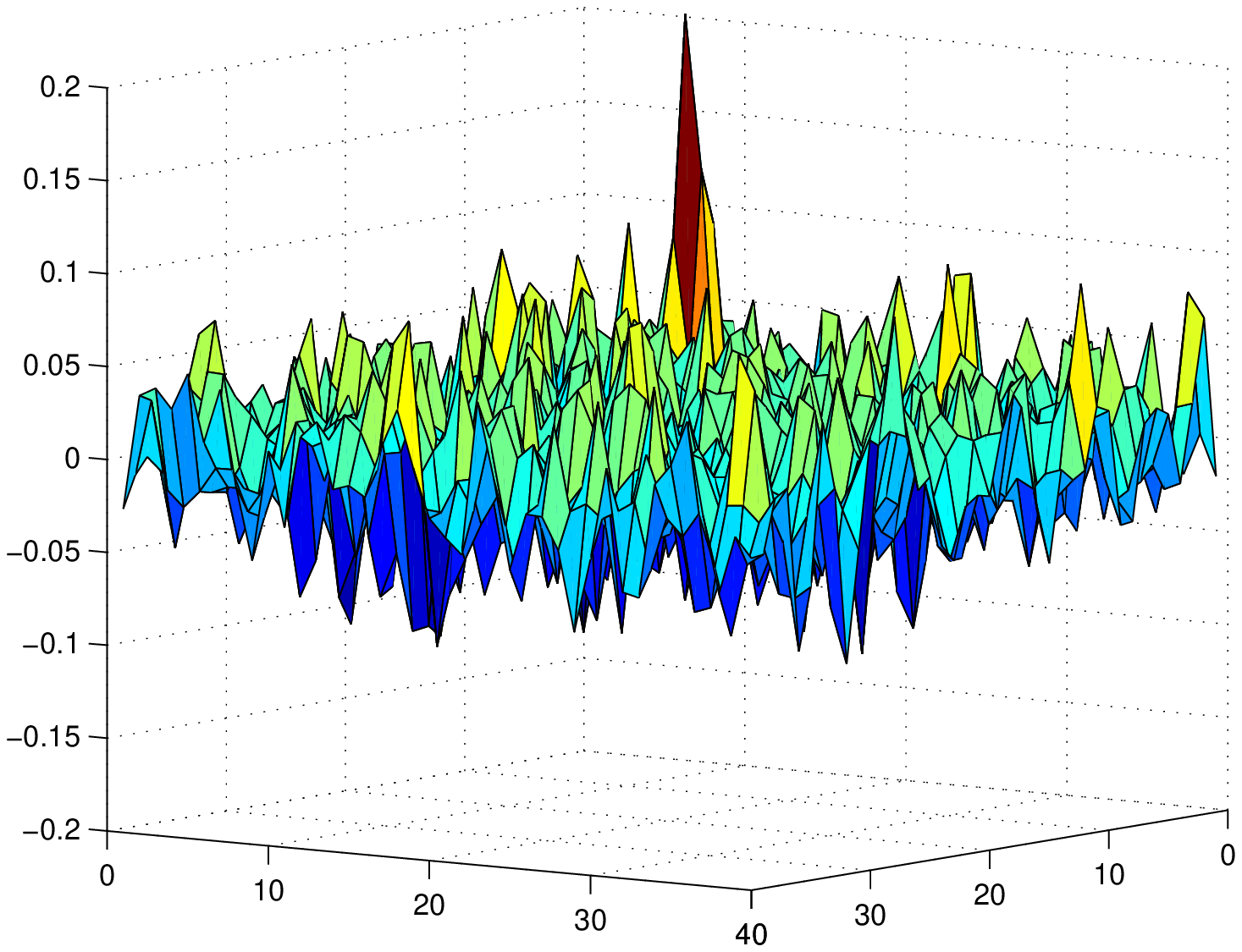}}\\
  \vspace{0in}
  \subfigure[$CP^3$, $\beta=1.6$, $Q=0$]{
    \label{fig:cp3_b16}
    \includegraphics[width=0.40\textwidth]{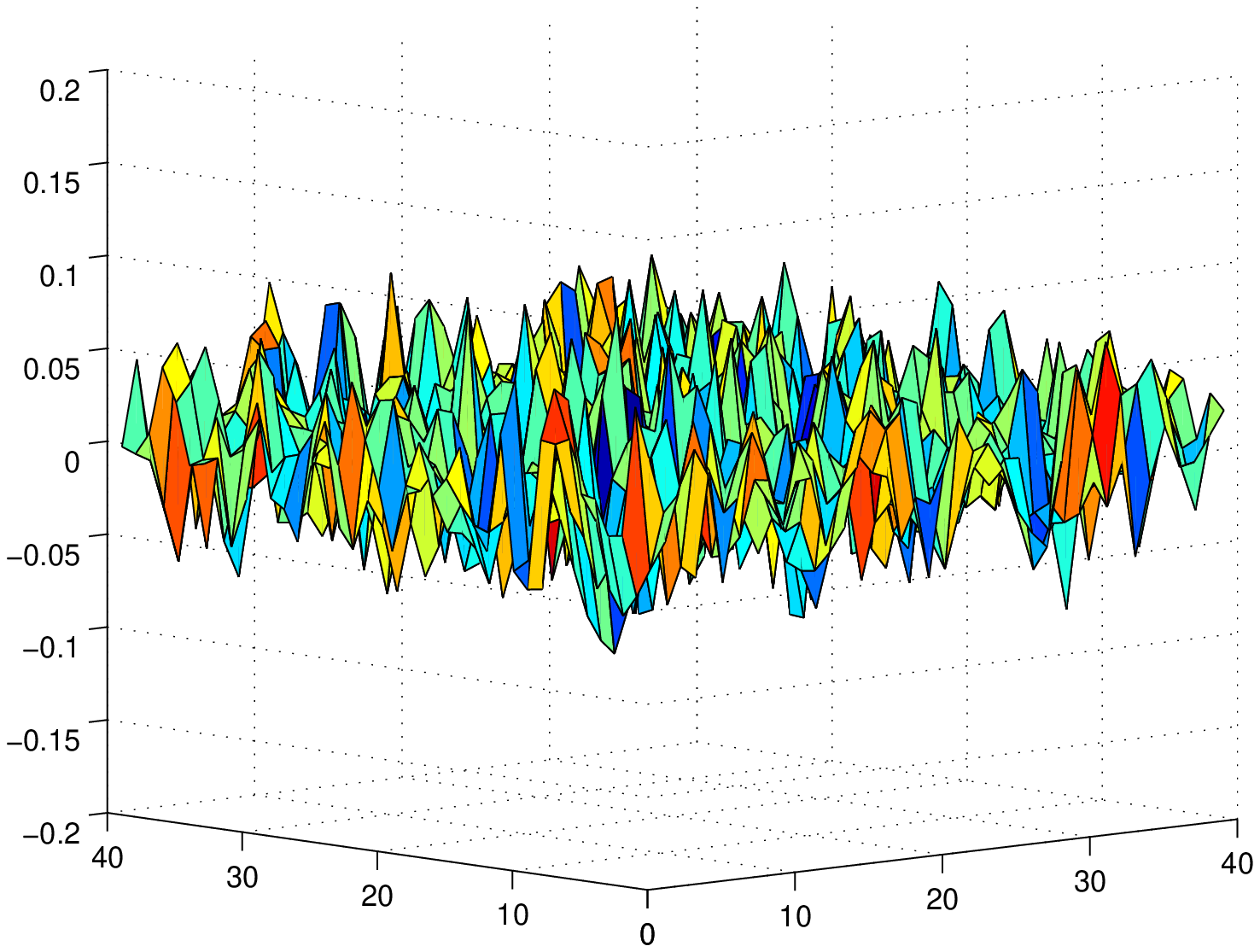}}
  \hspace{0in}
  \subfigure[$CP^9$, $\beta=0.9$, $Q=-1.0$]{
    \label{fig:cp9_b09}
    \includegraphics[width=0.40\textwidth]{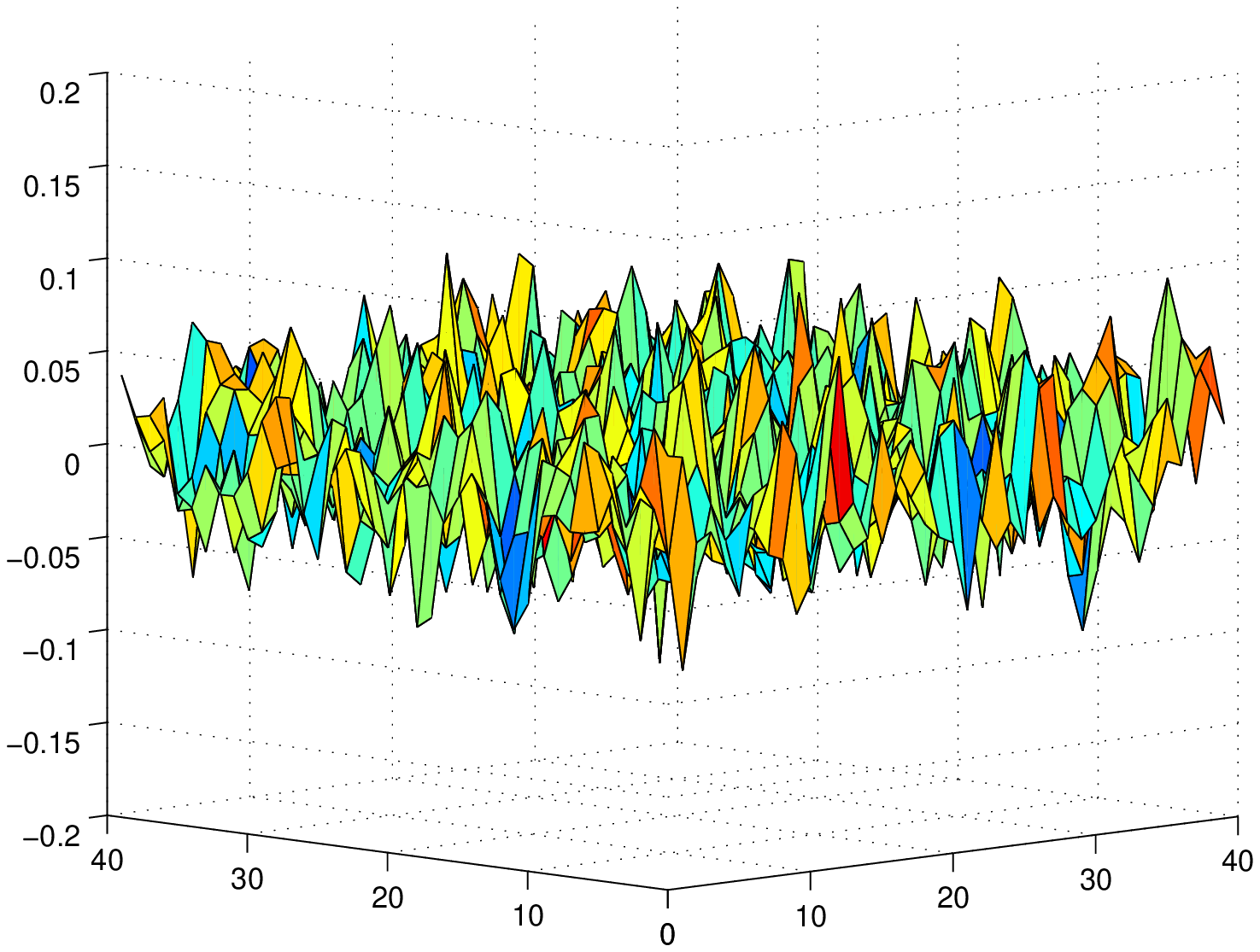}}
  \caption{\label{fig:small_instantons} 
(a) A peak structure is clearly seen in a 
$CP^1$ configuration with $\beta=2.5$ and $Q=1.0$. (b) No peak structures in 
$CP^1$ with $\beta=2.5$ and $Q=0$. 
(c) Two negative peak structures in $CP^1$ with $\beta=1.6$ and 
$Q=-2.0$.  (d) A peak structure in $CP^2$ with 
$\beta=1.8$ and $Q=1.0$. (e) $CP^3$ 
with $\beta=1.6$ and $Q=0$ shows no peak structure. (f) $CP^9$ with 
$\beta=0.9$ and $Q=-1.0$ shows no peak structure. }
\end{figure}

Even a qualitative survey of overlap topological charge distributions in the Monte Carlo configurations
is sufficient to identify small instantons as the origin of the divergence of $\chi_t$ in $CP^1$ and $CP^2$.
To emphasize this point, we will first present a few typical configurations. Small instantons
are particularly prominent and easy to identify at large values of $\beta$, where the 
meson correlation length $\mu^{-1}$ is large. 
Interestingly, the finite volume effects on the fluctuation of the global topological charge $Q$
are completely different for $CP^1$ and $CP^2$ than they are for the $N\geq 4$ models.
For $CP^3$ and higher, the fluctuation of $Q$ to nonzero integer values ``freezes out''
at large values of $\beta$ for a lattice with a fixed number of sites. When the correlation length becomes comparable to the box size,
$Q$ stops fluctuating altogether. In marked contrast, for example, $CP^1$ on a 40x40 lattice
continues to fluctuate to nonzero $Q$ well above $\beta=1.6$, where the correlation length $\mu^{-1}$ is greater than 40.
This is apparently due to the fact that the small instantons remain of order lattice spacing in size, even
when the correlation length becomes large. They are therefore not very susceptible to being squeezed out
by finite volume effects at the scale of the correlation length. 
In Fig.~\ref{fig:small_instantons} we show several typical configurations. Fig.~\ref{fig:cp1_b25_Q1} is a $CP^1$ 
configuration with $\beta=2.5$ and $Q=1.0$. It clearly shows a positively charged peak 
which is distinctly different from the background quantum fluctuations. We will refer to these objects, which are
easily identified in the large $\beta$ regime, as ``peak structures.'' As we discuss below, the amount of charge contained in 
a typical peak structure is approximately one unit, so they clearly exhibit local quantization of topological
charge.  Furthermore, for large values of $\beta$, fluctuations of the global charge $Q$ by a unit during the Monte Carlo run
are always accompanied by the appearance or disappearance of one of these peak structures.  
For comparison, we also show a $CP^1$ configuration with 
$\beta=2.5$ and $Q=0$ in Fig.~\ref{fig:cp1_b25_Q0}, where no peak structure can be seen. 
Fig.~\ref{fig:cp1_b16} shows two negative peak structures in a $CP^1$ configuration with 
$\beta=1.6$ and $Q=-2.0$. Because of the smaller value of $\beta$, 
the background quantum fluctuations are seen to be 
larger than in the previous plots. The $CP^2$ configuration in Fig.~\ref{fig:cp2_b18} 
also shows a peak structure with $\beta=1.8$ and $Q=1.0$. However, at comparable correlation 
lengths, the $CP^3$ and $CP^9$ configurations contain no prominent peaks and always have zero 
total topological charge on a $40\times40$ lattice. Fig.~\ref{fig:cp3_b16} shows 
a $CP^3$ configuration with $\beta=1.6$ and $Q=0$, where no peak structure can be seen. 
Finally, Fig.~\ref{fig:cp9_b09} shows a $CP^9$ configuration with 
$\beta=0.9$ and $Q=-1.0$. Here, in spite of the fact that Q is nonzero, there is no sign of an integer charged
peak structure. In fact, for $CP^5$ and $CP^9$, there were no integer charged peak structures in any of the
configurations studied. The case of $CP^3$ is interesting because, although there is some indication of 
small instantons (see below), they are not as dominant as in $CP^1$ and $CP^2$. Note that the one-instanton
action of $\epsilon=14.1$ for $CP^3$ is slightly above the instanton ``melting point'' of $4\pi=12.57\ldots$ 
predicted in the dilute gas approximation. In the case of $CP^3$, the small instantons do not lead to 
any observable anomalous scaling of $\chi_t$ because their action is very close to $4\pi$.  

To add further support for the assertion that the peak structures in $CP^1$ and $CP^2$ are indeed small instantons, we measured 
the topological charge inside the highest peak structure in each field configuration. As we will show in the
following discussion, a good effective definition of the topological charge within a peak is given by the sum of 
the charges within a radius $\leq 2$ of the maximum-charge site. To illustrate the quantization of the charge
in the peak structures, we analyze, for several values of $\beta$ and $N$, 
a large sample of configurations which have a global topological charge of
$Q=\pm1$. For each configuration we calculate the amount of charge within the highest peak structure. The
histograms show the number of configurations which have a charge within a given range for the highest structure.
As clearly shown, the $CP^5$ and $CP^9$ 
models exhibit very different histograms from $CP^1$, $CP^2$ and $CP^3$. The $CP^5$ and $CP^9$ configurations 
typically have much less than a unit of topological charge in the highest peak structure, 
while in $CP^1$, $CP^2$ and $CP^3$, the majority of $Q=\pm1$ configurations have a structure which contains 
approximately a unit of topological charge. This clearly confirms that the peak structures we observed in 
Fig.~\ref{fig:small_instantons} are in fact small instantons with a radius of $<2$ lattice spacings. By contrast,
the topological structures in higher $\CP$ models exhibit no tendency toward local quantization, consistent with 
expectations from the domain-wall picture of topological charge excitations in these models.

\begin{figure}
  \centering
  \subfigure[$CP^1$, $\beta=1.8$]{
    \label{fig:qstruct_cp1_b18}
    \includegraphics[width=0.30\textwidth]{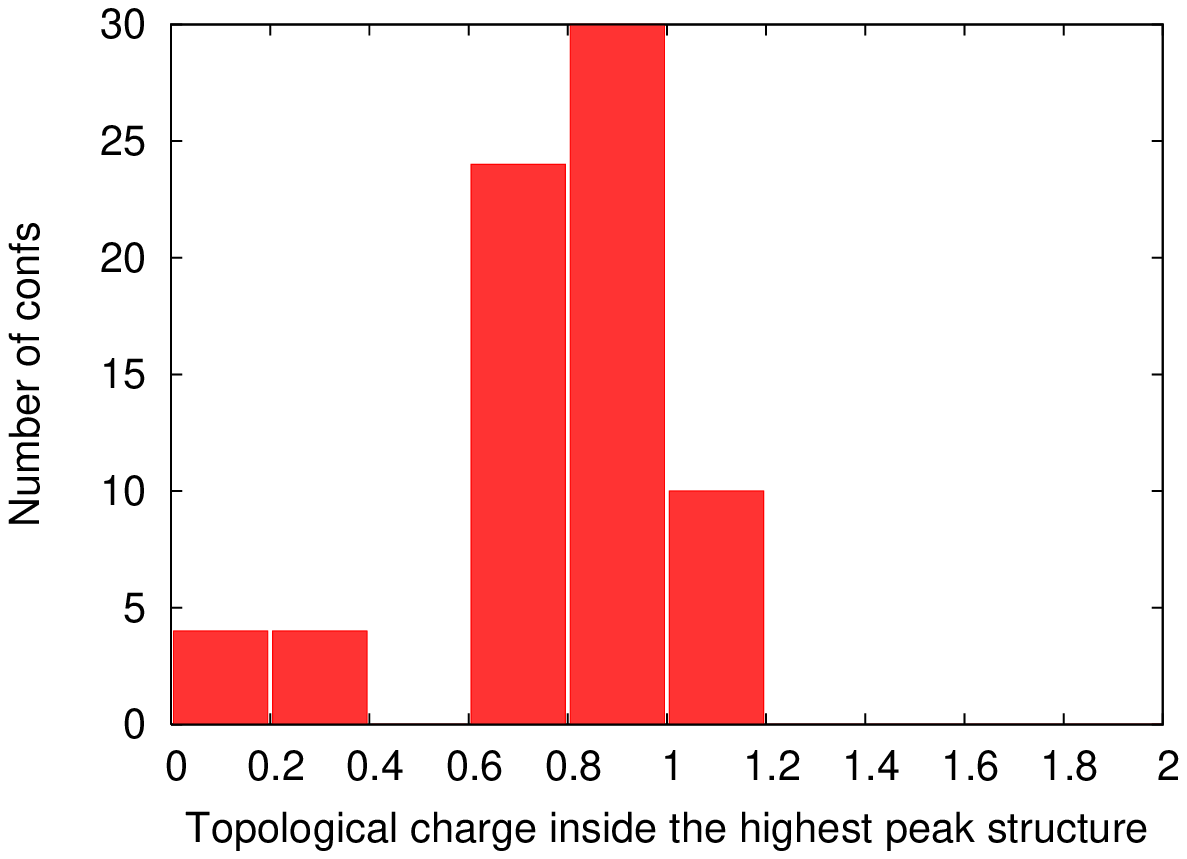}}
  \subfigure[$CP^1$, $\beta=2.0$]{
    \label{fig:qstruct_cp1_b20}
    \includegraphics[width=0.30\textwidth]{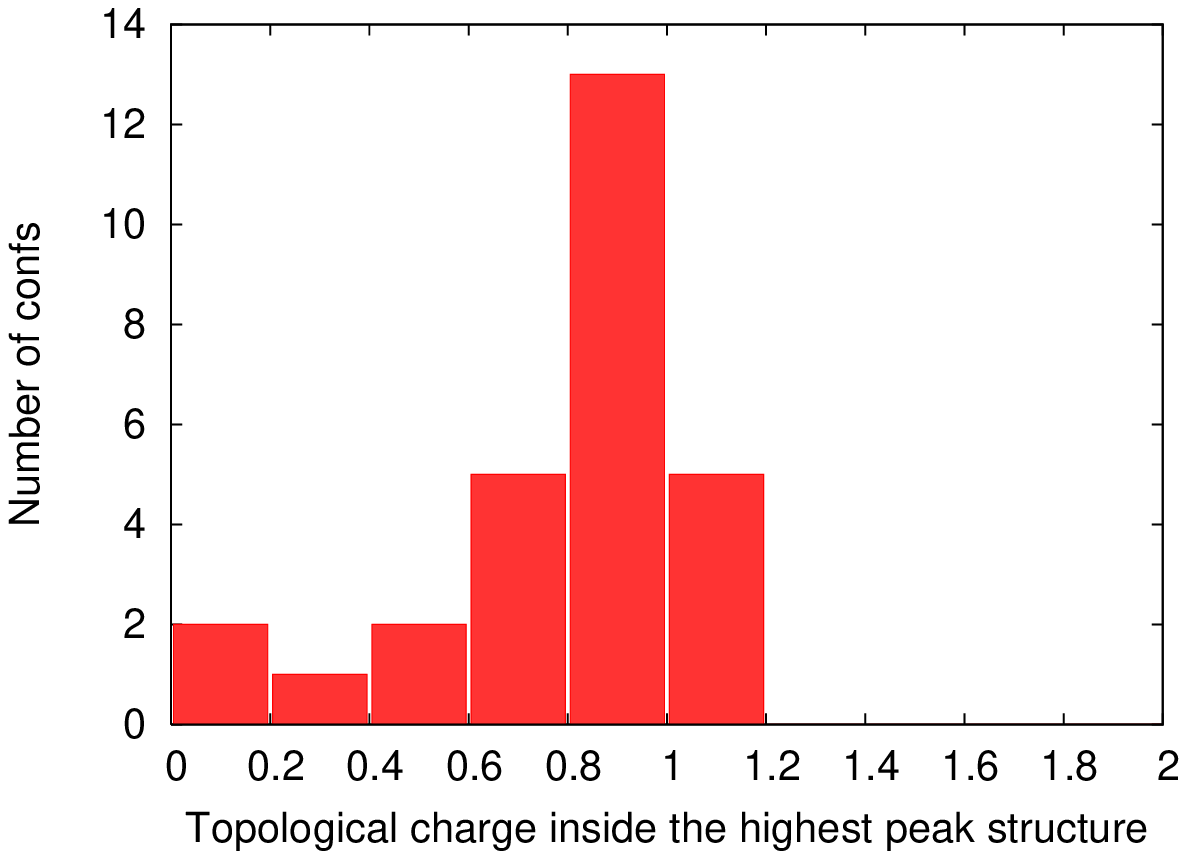}}
  \subfigure[$CP^2$, $\beta=1.6$]{
    \label{fig:qstruct_cp2_b16}
    \includegraphics[width=0.30\textwidth]{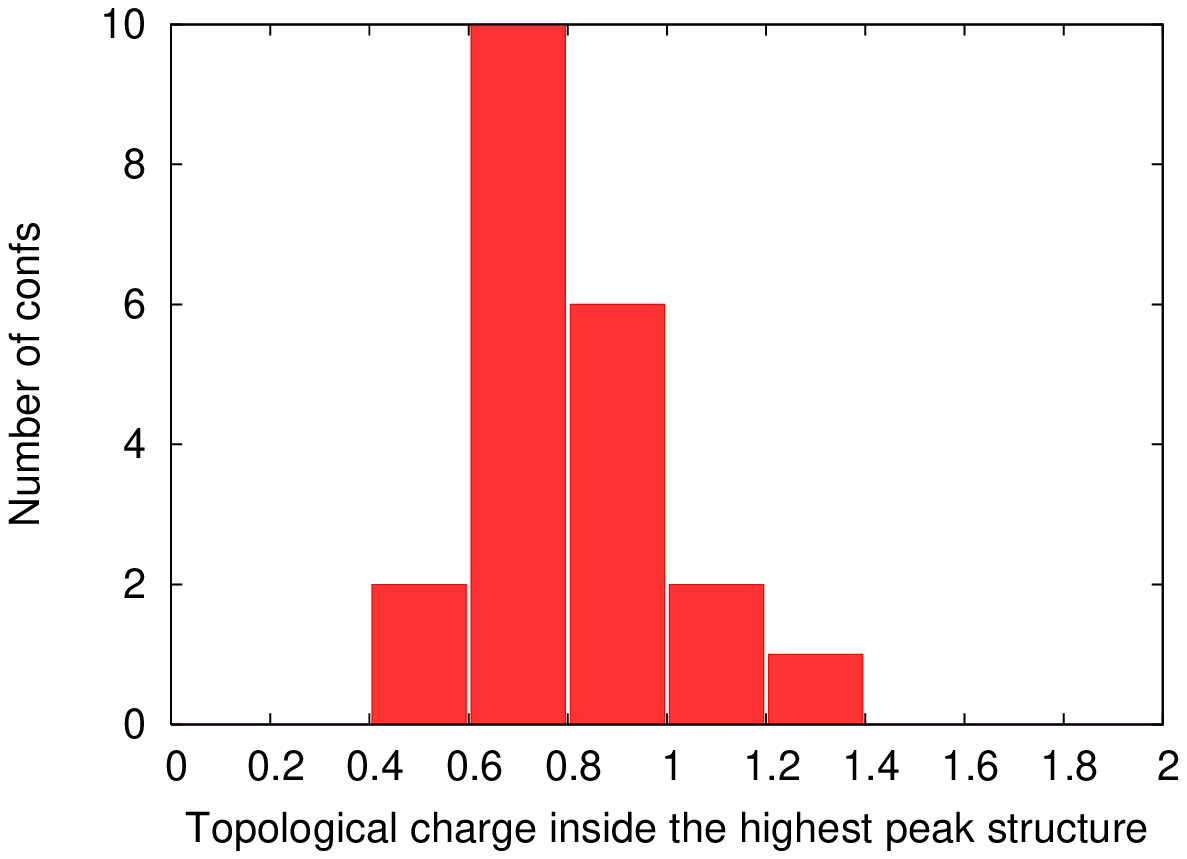}}\\
  \subfigure[$CP^3$, $\beta=1.2$]{
    \label{fig:qstruct_cp3_b12}
    \includegraphics[width=0.30\textwidth]{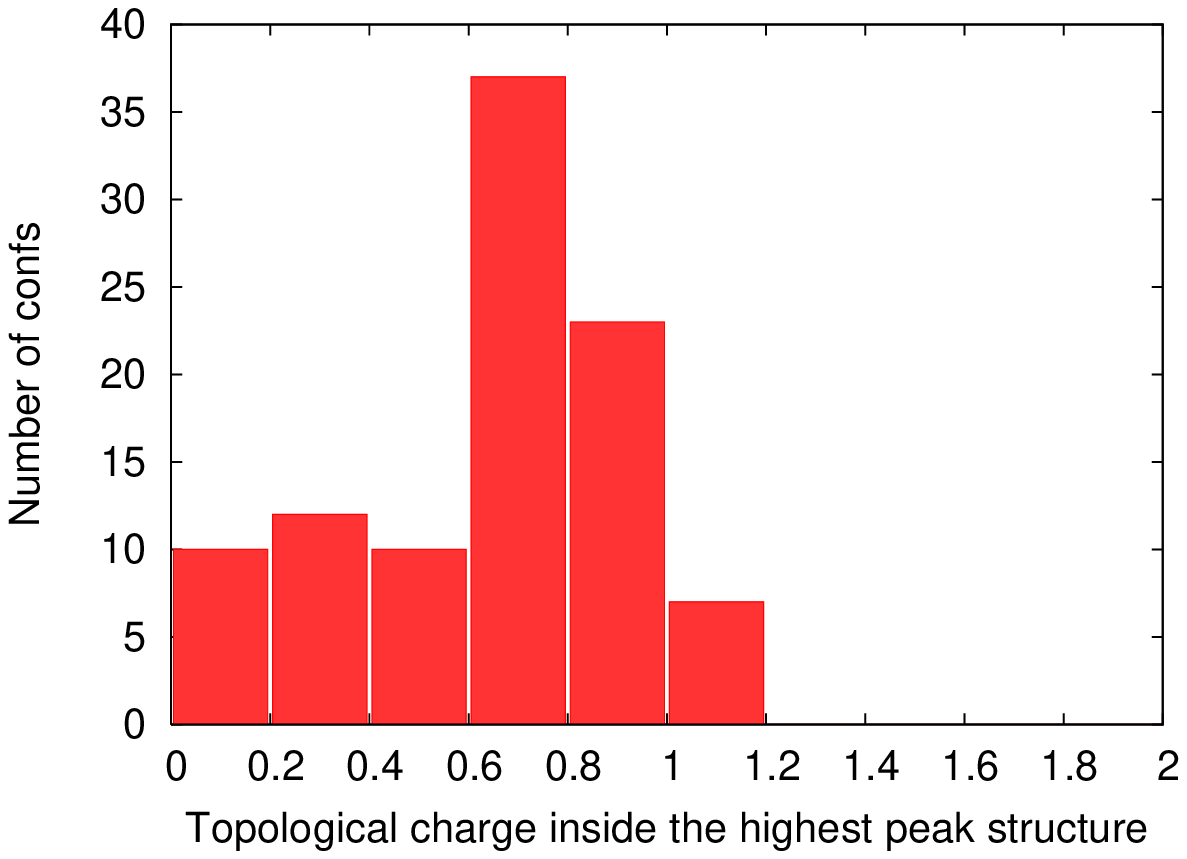}}
  \subfigure[$CP^5$, $\beta=1.0$]{
    \label{fig:qstruct_cp5_b10} 
    \includegraphics[width=0.30\textwidth]{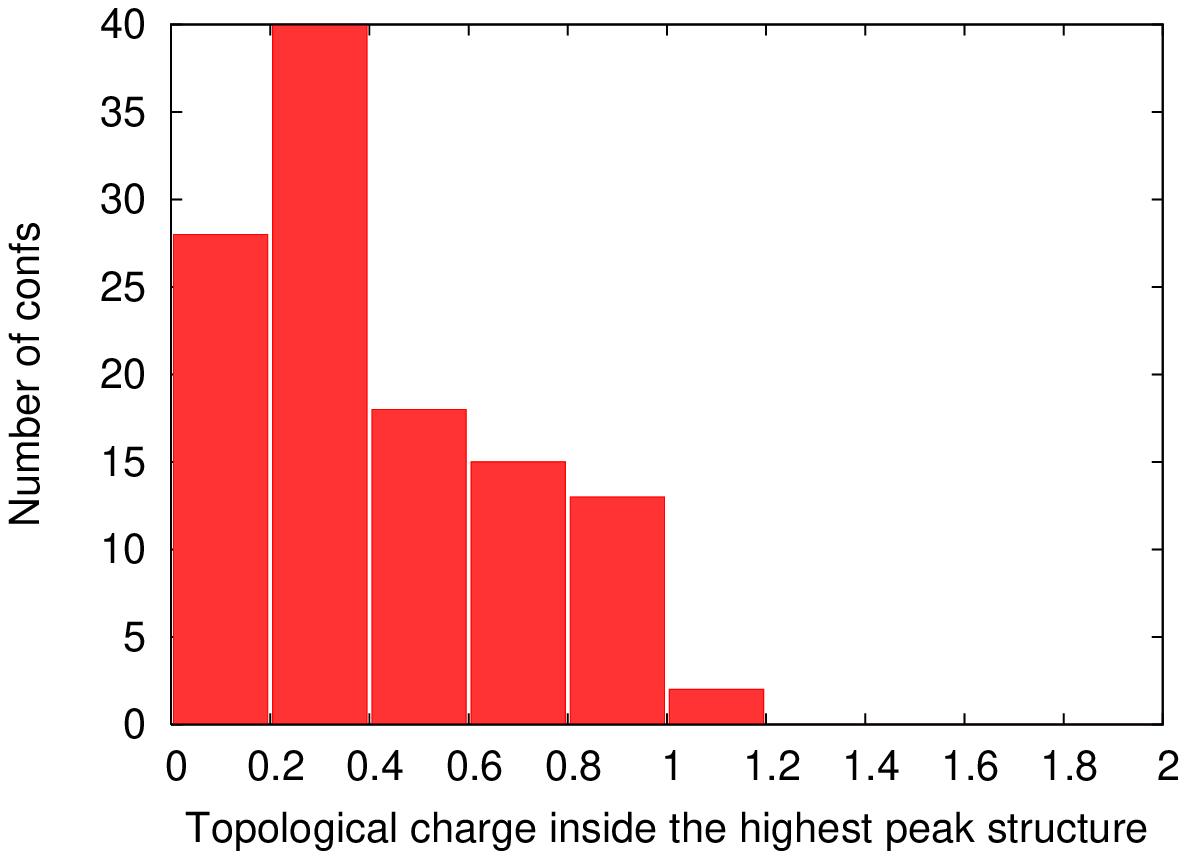}}
  \subfigure[$CP^9$, $\beta=0.8$]{
    \label{fig:qstruct_cp9_b08}
    \includegraphics[width=0.30\textwidth]{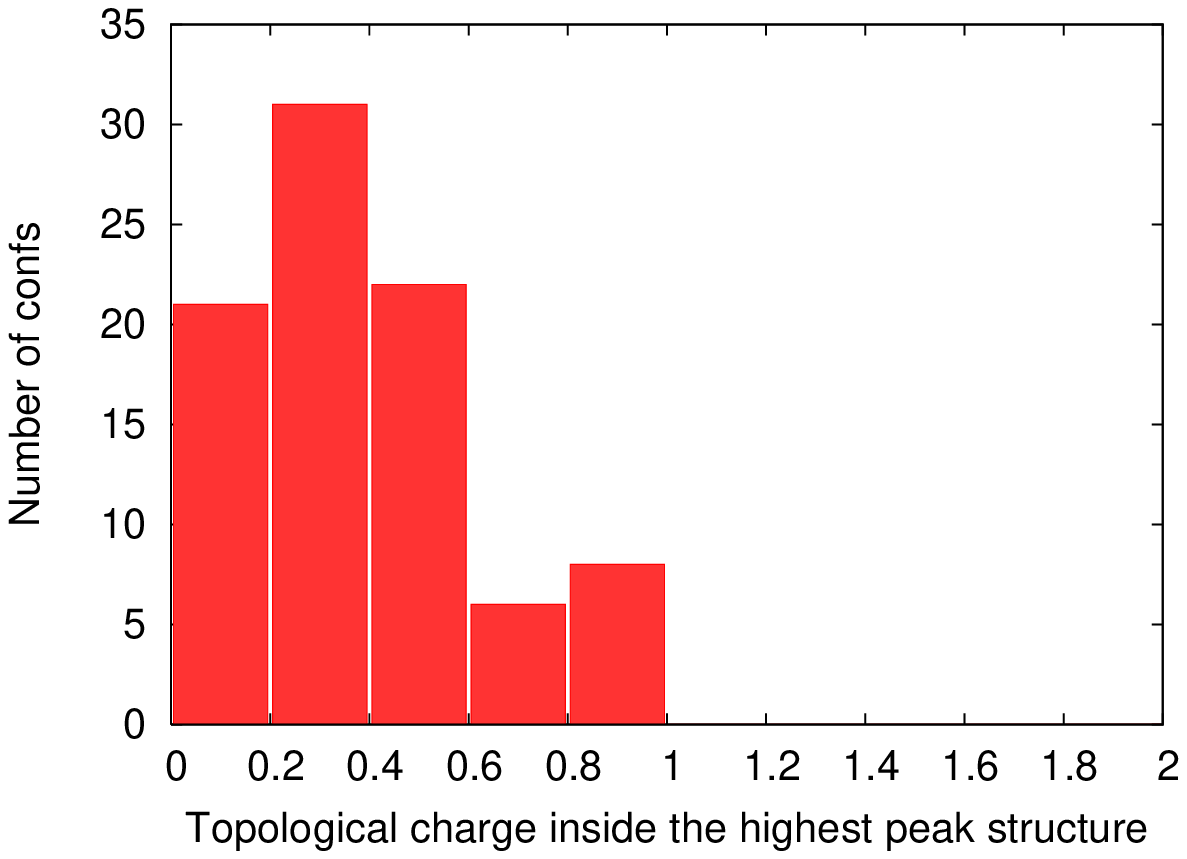}}
  \caption{\label{fig:structures} Histogram of the number of configurations with respect to the topological 
charge inside the highest peak structure}
\end{figure}

\begin{figure}
  \centering
  \subfigure[$q(r)$ and $Q(r)$ vs. $r$]{
    \label{fig:size_of_instanton}
    \includegraphics[width=0.45\textwidth]{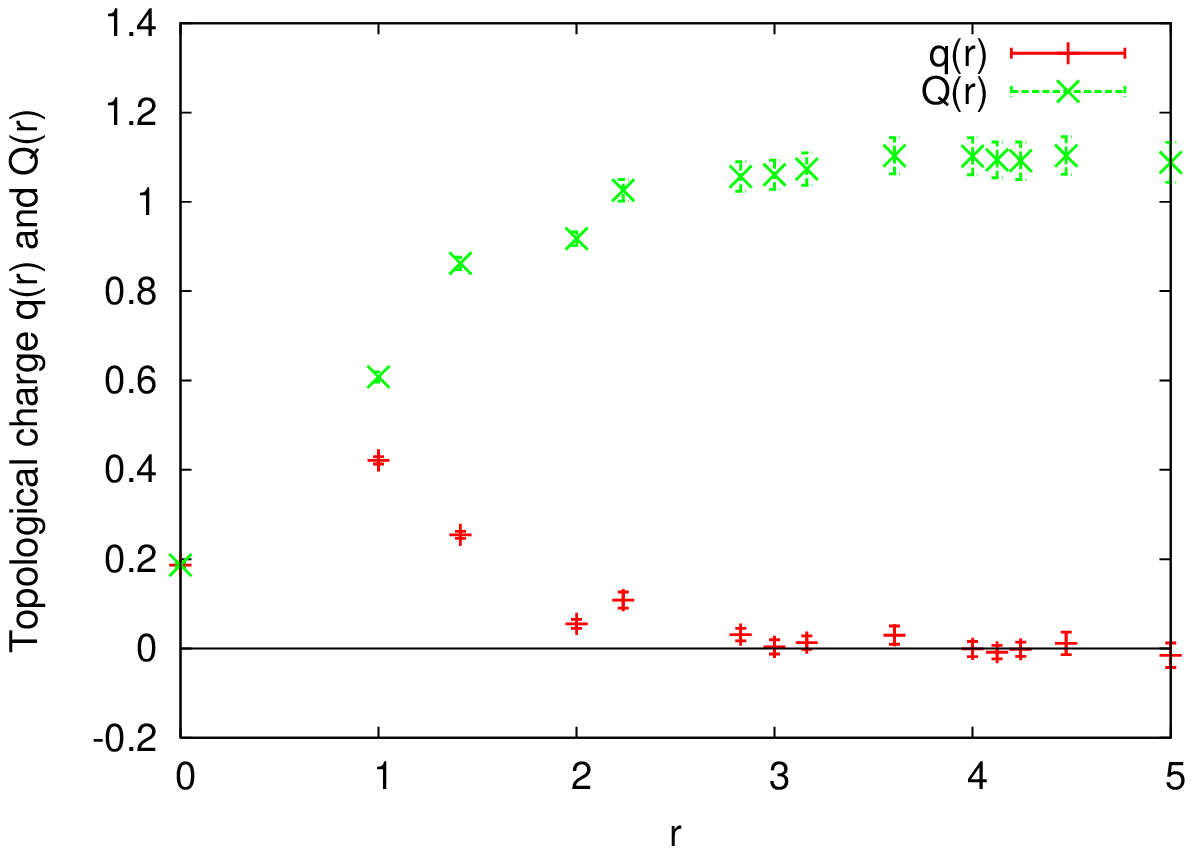}}
  \hspace{.2in}
  \subfigure[The action step]{
    \label{fig:action_step}
    \includegraphics[width=0.45\textwidth]{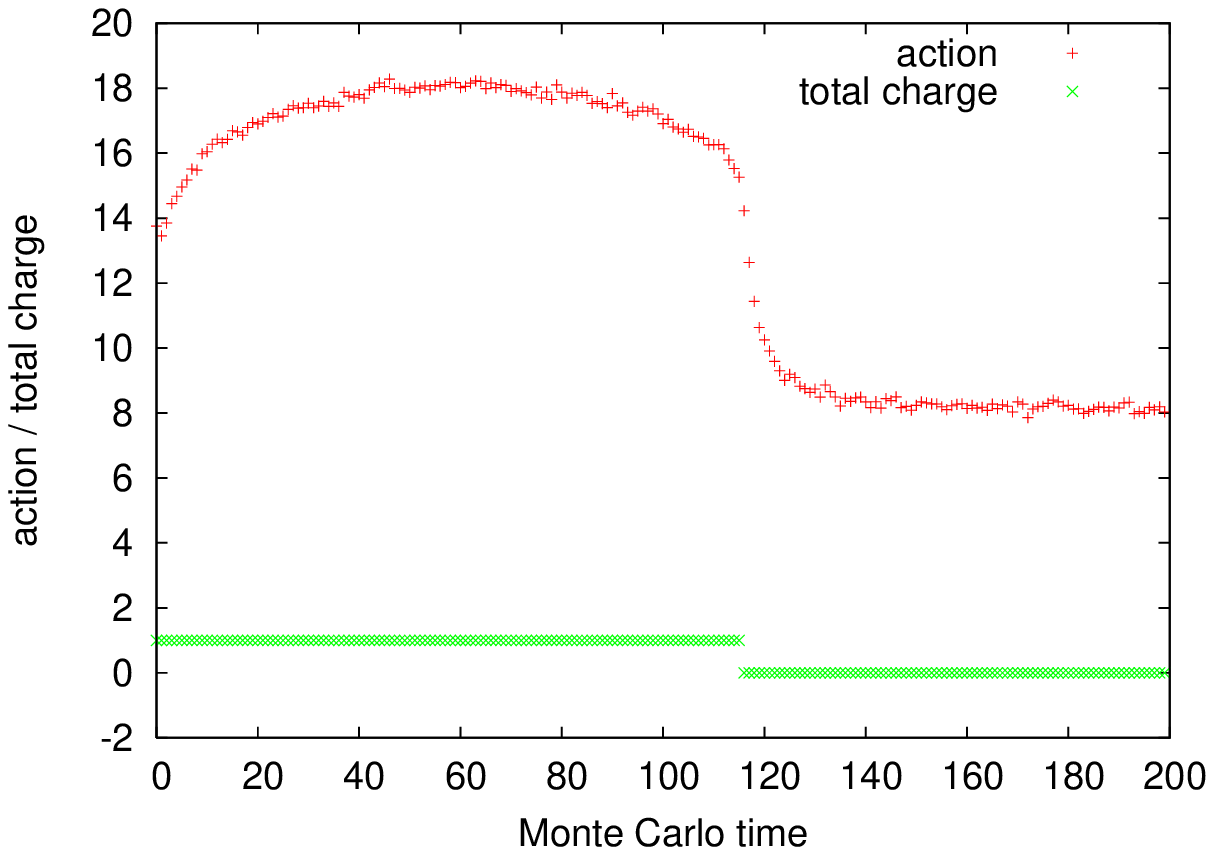}}
  \caption{\label{fig:radius_action} (a) The $r$ dependence of the topological charge density $q(r)$ and 
the total topological charge $Q(r)$ within radius $r$; (b) The gauge field action steps down when the 
$CP^1$ configuration tunnels from $Q=1$ into $Q=0$ sector. The upper curve is the action and the lower curve 
is the charge $Q$.}
\end{figure}

To estimate the approximate radial size of the peak structures, we considered an ensemble of $CP^1$ configurations 
on a 40x40 lattice at $\beta=1.8$, where we have $72$ $Q=\pm 1$ configurations out of $400$. 
In each configuration, we started with the highest peak, 
then measured the topological charge density $q(r)$ and the integrated topological charge $Q(r)$ within radius $r$. 
Fig.~\ref{fig:size_of_instanton} shows the $r$ dependence of $q(r)$ and $Q(r)$ with error bars. We see from the 
graph that $q(r)$ drops to zero at about $r\approx 2$, while at the same time $Q(r)$ goes up to $\approx 1.0$. For this reason, we 
define the peak structure as occupying the lattice sites within a radius of two lattice spacings from the peak.  
We also observe from the graph that the $q(r)$ fluctuates around zero when $r>2.0$ and the fluctuation is comparable 
to error bars. This confirms that the radial size of the peak structures we observed in Fig.~\ref{fig:small_instantons} 
is indeed roughly 1 to 2 lattice spacings. All of these properties fit precisely the description of small
instantons, which arise on the lattice as minimum action configurations in the $Q=\pm1$ sector. 

In order to compare topological susceptibility measurements with the predictions of a dilute gas of small instantons,
we want to determine the action of a single small instanton by 
measuring the minimal action in the $Q=1$ sector for various $\CP$ models. 
The value of the minimal action for a small instanton determines whether they will dominate the 
topological fluctuations in the continuum 
limit. For $CP^1$ and $CP^2$ the value of the minimal action also determines the form of the divergence of $\chi_t$
in the continuum limit. We constructed a small instanton configuration on the 
lattice, which consists of three vertical links of $e^{i\eta}$, $e^{i\pi}$, and $e^{-i\eta}$ respectively  
on three adjacent sites on a horizontal row, while all other links are set to one. 
Here the value of $\eta$ is taken at the boundary between the $Q=1$ and $Q=0$ sectors, as determined by the
overlap $Q$. This gives, $\eta = .0873(1)$.
We then annealed this configuration (by running the Monte Carlo on the $z$ fields at large $\beta$ with fixed 
gauge links) and measured its action. The minimal actions we obtained are: $\epsilon = 6.76$ 
for $CP^1$, $\epsilon = 10.41$ for $CP^2$ and $\epsilon = 14.08$ for $CP^3$. For $CP^1$ this is close to L\"{u}scher's 
result $\epsilon = 6.69$ \cite{Luscher82}. Since our lattice action and definition of topological charge
differ from the $O(3)$ model formulation, 
exact agreement between our measured $\epsilon$ and that of Ref.\cite{Luscher82} is not expected. We also looked at 
how the lattice action changes when we let a $CP^1$ configuration tunnel from the $Q=1$ sector into the $Q=0$ sector,
starting with a Monte Carlo generated $Q=1$ configuration and cooling it by running at
a very large $\beta$. An action step is expected which should be close to the minimal action in the $Q=1$ 
sector. Fig.~\ref{fig:action_step} clearly shows the action step when the tunneling occurs. The action step is 
measured to be about $7$, which is very close to the one-instanton action. 

\begin{figure}[t]
  \centering
  \subfigure[$CP^1$]{
    \label{fig:cp1_defect}
    \includegraphics[width=0.45\textwidth]{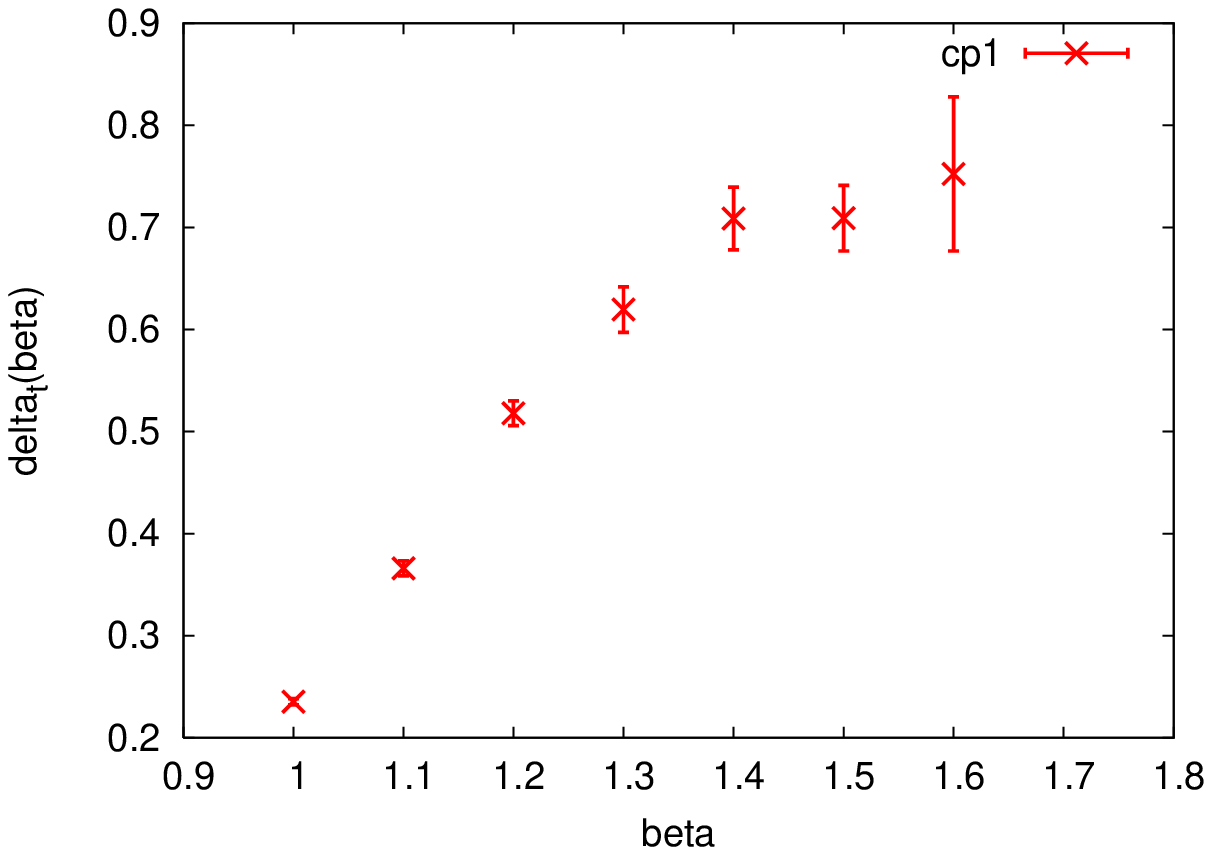}}
  \hspace{.2in}
  \subfigure[$CP^2$]{
    \label{fig:cp2_defect}
    \includegraphics[width=0.45\textwidth]{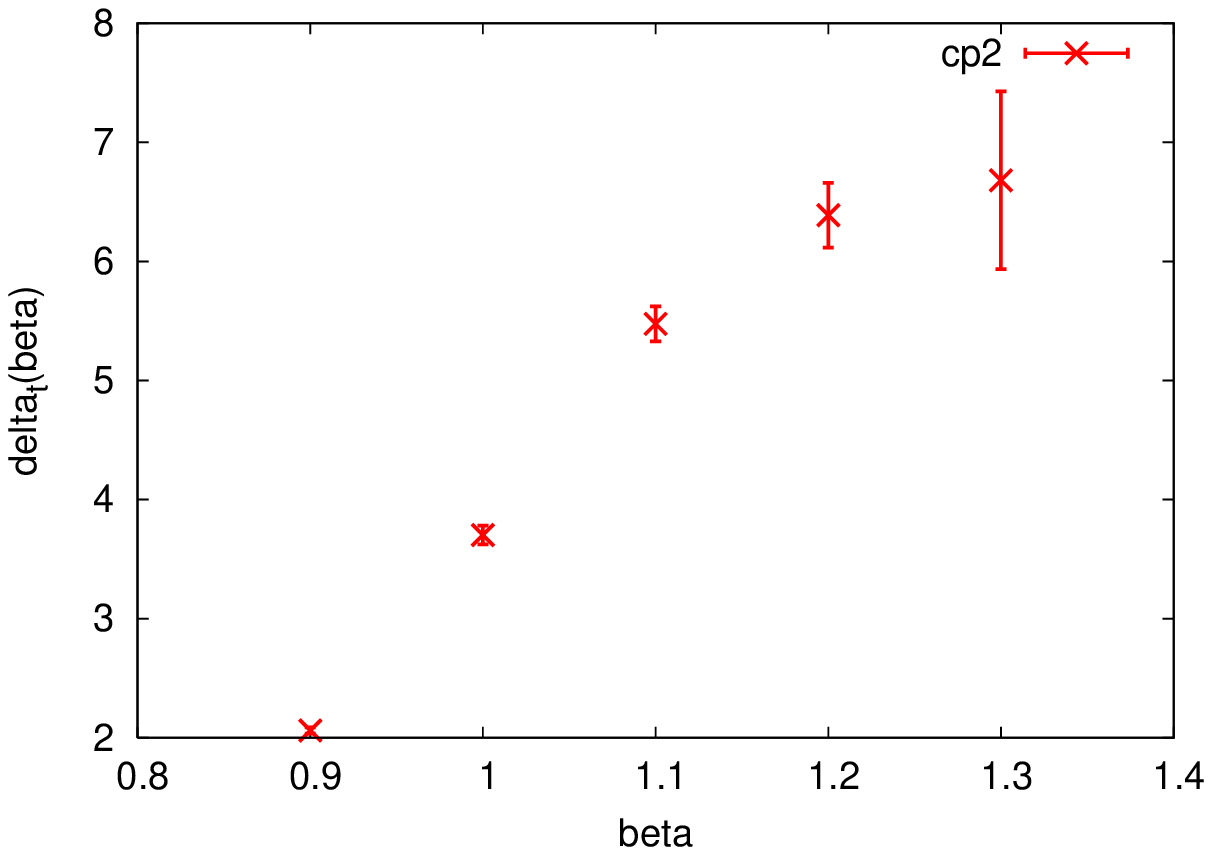}}
  \vspace{.2in}
  \subfigure[$CP^5$]{
    \label{fig:cp5_defect}
    \includegraphics[width=0.45\textwidth]{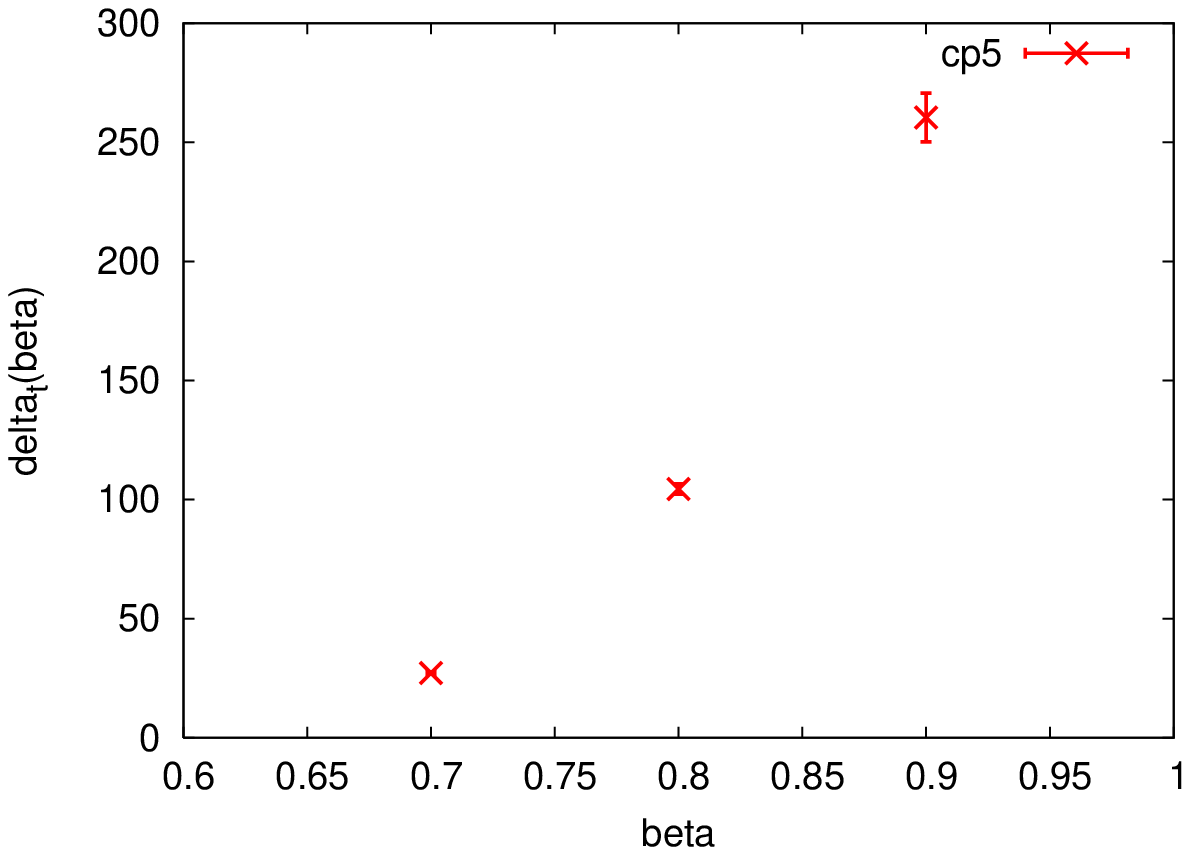}}
  \hspace{.2in}
  \subfigure[$CP^9$]{
    \label{fig:cp9_defect}
    \includegraphics[width=0.45\textwidth]{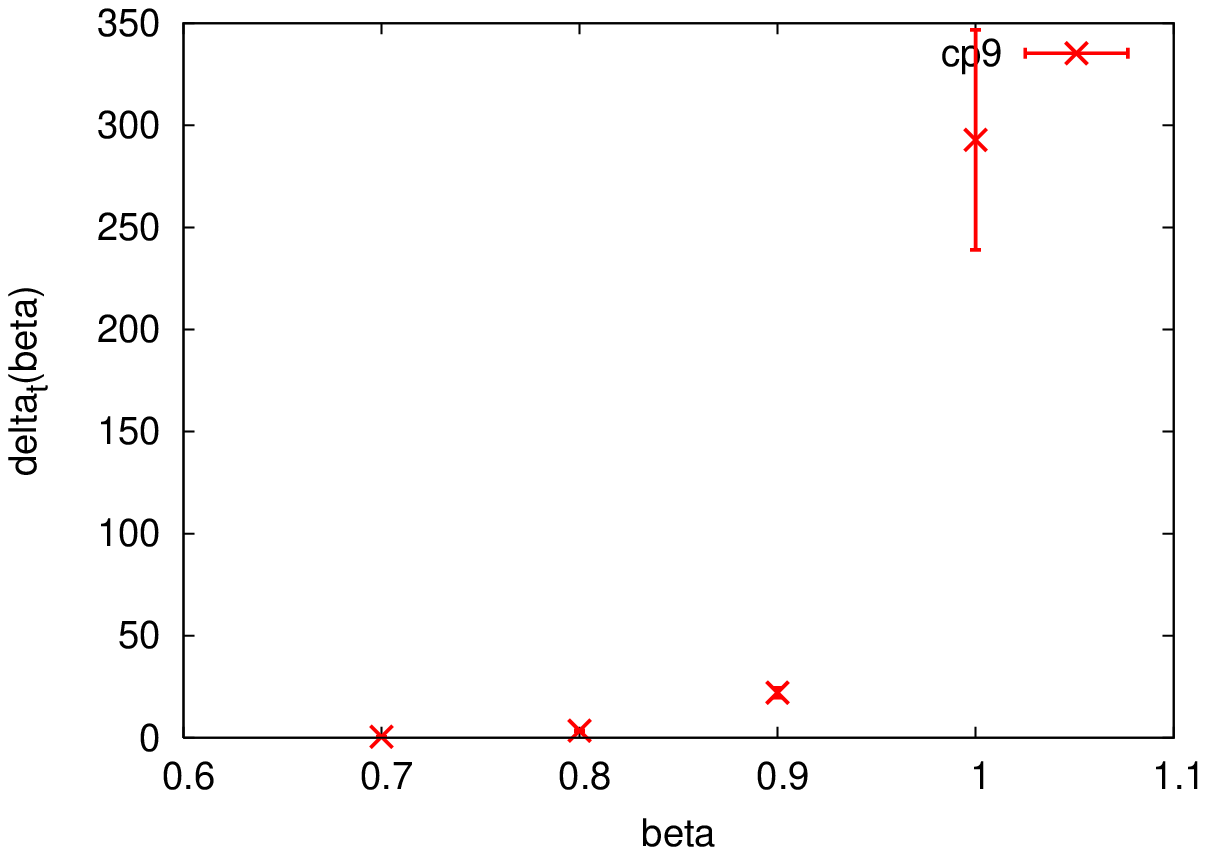}}
  \caption{\label{fig:cpn_defect} Plot of the coefficient $\delta_t(\beta)$ defined in Eqn.~\ref{eqn:delta_t}.}
\end{figure}

Some insight into the dynamics of small instantons in $CP^1$ and $CP^2$ can be obtained by comparing the
measured topological susceptibility with the prediction of a dilute instanton gas calculation. 
In the dilute gas approximation, the topological susceptibility is determined by the one-instanton
action $\epsilon$, and behaves like
\begin{equation}
\label{eqn:chi_t_dg2}
\chi_{t}^{d.g.} \propto \beta^{-1} e^{-\beta\epsilon}, \quad \quad \left(\beta \to \infty \right)
\end{equation}
To test how well the dilute gas approximation works, we plot the quantity
\begin{equation}
\label{eqn:delta_t}
\delta_{t}(\beta)= \beta e^{\beta\epsilon} \chi_t(\beta)
\end{equation}
against $\beta$ in Fig.~\ref{fig:cpn_defect}, up to an irrelevant overall factor. 
Fig.~\ref{fig:cp1_defect} and \ref{fig:cp2_defect} show that $\delta_t$ becomes roughly constant 
at large $\beta$ for $CP^1$ and $CP^2$, confirming the dilute gas calculation. 
However, $\delta_t$ is a rapidly increasing function of $\beta$ for $CP^5$ and $CP^9$,
showing that the expected contribution from instantons (\ref{eqn:chi_t_dg2}) is much
smaller than the measured topological susceptibility. This confirms the conclusion 
that small instantons are irrelevant to the determination of $\chi_t$ for $N>4$.

In the next section we discuss the mechanism of topological charge fluctuations in different 
$\CP$ models. We will argue that small instantons and membrane structures are the two types of lower dimensional
topological structures (zero- and one-dimensional respectively) that can appear in $\CP$ models.

\section {Mechanism of Topological Charge Fluctuations}
\label{sec:mechanism}
As we have shown in the previous section, small instantons dominate topological charge fluctuations 
in $CP^1$ and $CP^2$ and produce anomalous scaling behavior of $\chi_t$, while $CP^3$ and higher $\CP$ models
exhibit proper scaling behavior $\chi_t\sim \mu^2$. Thus, in the continuum limit, $CP^1$ and $CP^2$ have
infinite topological susceptibility. The original observation of this divergent susceptibility \cite{BergLuscher} was 
obtained with an ultralocal definition of topological charge in the O(3) sigma model formulation. 
The fact that the same divergent behavior of $\chi_t/\mu^2$ occurs when one uses the overlap construction
suggests that the divergence cannot be eliminated by a better lattice definition of topological charge, and that it
has some physical significance in the continuum limit. To gain a better understanding of this, it is useful to
consider the topological charge correlator in Euclidean 2-space,
\begin{equation}
\label{eq:TC}
G(x) = \langle q(x)q(0) \rangle
\end{equation}
The lattice studies discussed here and in Ref. \cite{Ahmad} provide clear evidence that, in the continuum limit,
this correlator is very short range (compared to the correlation length $\mu^{-1}$), so one might expect an
operator product expansion to provide insight.  
Here we consider an operator product expansion in the effective $U(1)$ gauge theory obtained by
integrating out the $z$-fields. Taking the large N solution as a guide, we expect that the charged $z$-particles acquire a ``constituent
quark mass'' $M$. At the same time, $z$-loop effects produce
a dynamically generated kinetic $F_{\mu\nu}^2$ term in the effective gauge action. This effect leads to a confining 
Coulomb potential between constituent $z$-particles. However, in the large $N$ limit, the confining potential is weak, 
corresponding to a string tension of order $M^2/N$. Thus, one can distinguish between a relatively long distance
scale $\sim \sqrt{N}M^{-1}$ associated with confinement, and a shorter distance scale of order $M^{-1}$, which probes the substructure of the
constituent $z$ particles. By invoking the operator product expansion in the effective gauge theory
obtained by integrating out the $z$-fields, we are assuming that the distance scales relevant to topological
structure are characterized by the confinement scale. The large $N$ solution, discussed below, 
provides a good illustration of the order of limits involved here.
In the large $N$ approximation, after integrating
out the $z$ fields, the action $L(x)$ can be replaced by 
\begin{equation}
\label{eq:effective}
L(x) \rightarrow \frac{1}{4}F_{\mu\nu}^2
\end{equation}
where $F_{\mu\nu}$ is a rescaled field strength obtained by the replacement
\begin{equation}
\label{eq:rescaledA}
A_{\mu}\rightarrow \sqrt{\frac{12\pi M^2}{N}}A_{\mu}
\end{equation}

The two lowest dimension gauge invariant, Lorentz invariant local
operators that can appear in the OPE for (\ref{eq:TC}) are the identity and the squared field strength $F_{\mu\nu}^2 \equiv F^2$,
\begin{equation}
\label{eq:OPE}
q(x)q(0) \sim C_1(x) + C_2(x) F^2(0) + \ldots
\end{equation}
where $\ldots$ represents operators of dimension greater than 2. 
In the following discussion, it will also be useful to consider the corresponding OPE coefficients for the product of
two Chern-Simons currents. Defining
\begin{equation}
J_{\mu}^{CS}(x) = \frac{1}{2\pi}\epsilon_{\mu\nu}A_{\mu}(x)
\end{equation}
we write the OPE (ignoring possible gauge-dependent terms which do not contribute to (\ref{eq:OPE})),
\begin{equation}
\label{eq:OPE_CS}
J_{\mu}^{CS}(x)J_{\nu}^{CS}(0) \sim C_{1\mu\nu}(x) + C_{2\mu\nu}(x) F^2(0) + \ldots
\end{equation}
The Chern-Simons correlator 
\begin{equation}
G^{CS}_{\mu\nu}(x) =\langle J_{\mu}^{CS}(x)J_{\nu}^{CS}(0)\rangle
\end{equation}
is related to the topological charge correlator (\ref{eq:TC}) by
\begin{equation}
G(x) = -\partial_{\mu}\partial_{\nu}G^{CS}_{\mu\nu}(x)
\end{equation} 
In the free theory defined by (\ref{eq:effective}), the behavior of the OPE coefficients is
given by naive dimensional counting. For the topological charge OPE, this would give.
\begin{eqnarray}
C_1(x)\sim 1/x^4 \\
C_2(x)\sim 1/x^2
\end{eqnarray}
However, we will see in the following discussion that, although such power-law behavior is true in
the OPE for the Chern-Simons correlator, the corresponding OPE for the topological charge correlator ``collapses''
to pure contact terms of the form
\begin{eqnarray}
C_1(x) & \sim & \nabla^2\delta^2(x) \\
C_2(x) & \sim & \delta^2(x)
\end{eqnarray} 
As we argue below, this collapsing of the operator product expansion for $G(x)$ is a consequence of the lower dimensional 
character of the small instanton and domain wall excitations of topological charge.
The Fourier transformed OPE coefficients 
\begin{equation}
\tilde{C}_i(q) = \int d^2x C_i(x) e^{iq\cdot x}
\end{equation}
have large $q^2$ behavior
\begin{eqnarray}
\tilde{C}_1(q^2)\sim q^2 \\
\tilde{C}_2(q^2)\sim 1
\end{eqnarray}
The singular terms in the operator product expansion correspond to polynomial subtraction terms in the dispersion
integral representation of the two-point correlator. We consider the correlator in momentum space,
\begin{equation}
\label{eq:qcor}
\tilde{G}(q^2) = \int d^2x e^{iq\cdot x} G(x)
\end{equation}
Under general assumptions, the correlator can be written in a dispersion integral representation \cite{Seiler87,Seiler01},
\begin{equation}
\label{eq:dispersion}
\tilde{G}(q^2) = c_1q^2 + c_2 - \int_{M_0^2}^{\infty} d\lambda^2\frac{\rho(\lambda^2)}{q^2 + \lambda^2}
\end{equation}
(Note: This representation holds in two space-time dimensions. In four dimensions, the polynomial part can include
a $q^4$ term.) The first two terms represent contact terms associated with the short distance singularities of 
the OPE (\ref{eq:OPE}). The integral term represents the contribution of real intermediate states of invariant
mass $\lambda$. Here $M_0$ is the mass gap in the flavor singlet channel. 
The standard way of analyzing correlators on the lattice is to pick a time direction and Fourier
transform over the spatial direction with spatial momentum component $q_1=0$, i.e. 
\begin{equation}
\hat{G}(t) = \int dx_1 G(x_0=t,x_1)
\end{equation}
The dispersion representation then gives
\begin{equation}
\label{eq:corr}
\hat{G}(t) = -c_1 \delta^{''}(t) + c_2\delta(t) - \int_{M_0^2}^{\infty} \frac{d\lambda^2}{2\lambda}\rho(\lambda^2)e^{-\lambda |t|}
\end{equation}
The topological susceptibility is 
\begin{equation}
\chi_t = c_2 - \int_{M_0^2}^{\infty} \frac{d\lambda^2}{\lambda^2}\rho(\lambda^2)
\end{equation}
Note that the $c_1$ contact term arises from the identity operator term in the OPE, while the $c_2$ term
arises from the $F^2$ insertion term. The $c_1$ term does not contribute to $\chi_t$ and the $c_2$ term must be positive
and larger in magnitude than the negative contribution from the integral over real intermediate states in
order for $\chi_t$ to be positive.

A detailed numerical analysis of the spectral and scaling properties of the $q(x)$ correlator 
will be reported elsewhere. Here we will quote and utilize some of the main properties of
the correlator which emerge from this numerical study. The results are found to be consistent with the assumption that
the correlator is reasonably well-described by contact terms alone, without the need for an intermediate state
term. We should emphasize that a description of the correlator as purely contact terms does not mean that real
propagating intermediate states do not contribute, but only that those states are heavy compared to the $q^2$
values we are probing on the lattice when we calculate the two-point correlator. 
The effect of heavy intermediate states can be absorbed into the contact
terms via
\begin{eqnarray}
c_1 & \rightarrow & c_1 + \int_{M_0^2}^{\infty}\frac{d\lambda^2}{\lambda^4}\rho(\lambda^2)\\
c_2 & \rightarrow & c_2 - \int_{M_0^2}^{\infty}\frac{d\lambda^2}{\lambda^2}\rho(\lambda^2)
\end{eqnarray}

The large N solution \cite{Witten79,Campostrini} provides an explicit example which illustrates 
how the pure contact term approximation to $G$ arises.
We consider the gauge field vacuum polarization tensor (the inverse of the 
$\langle A_{\mu}A_{\nu}\rangle$ correlator) in momentum space, 
\begin{equation}
\Delta_{\mu\nu}^{-1} \equiv \left(\delta_{\mu\nu}-\frac{q_{\mu}q_{\nu}}{q^2}\right)\left(\Delta(q^2)\right)^{-1}
\end{equation}
Since $2\pi q(x)$ is just the curl of $A_{\mu}$, 
\begin{equation}
q(x) = \frac{1}{2\pi}\epsilon_{\mu\nu}\partial_{\mu}A_{\nu}
\end{equation}
the gauge correlator is related to the topological charge
correlator by
\begin{equation}
4\pi^2\tilde{G}_{\mu\nu}^{CS}(q) = \frac{q_{\mu}q_{\nu}}{q^2}\Delta(q^2)
\end{equation}
or
\begin{equation}
4\pi^2 \tilde{G}(q^2) = q^2\Delta(q^2)
\end{equation}

\begin{figure}
  \centering
  \includegraphics[width=0.4\textwidth, angle=0]{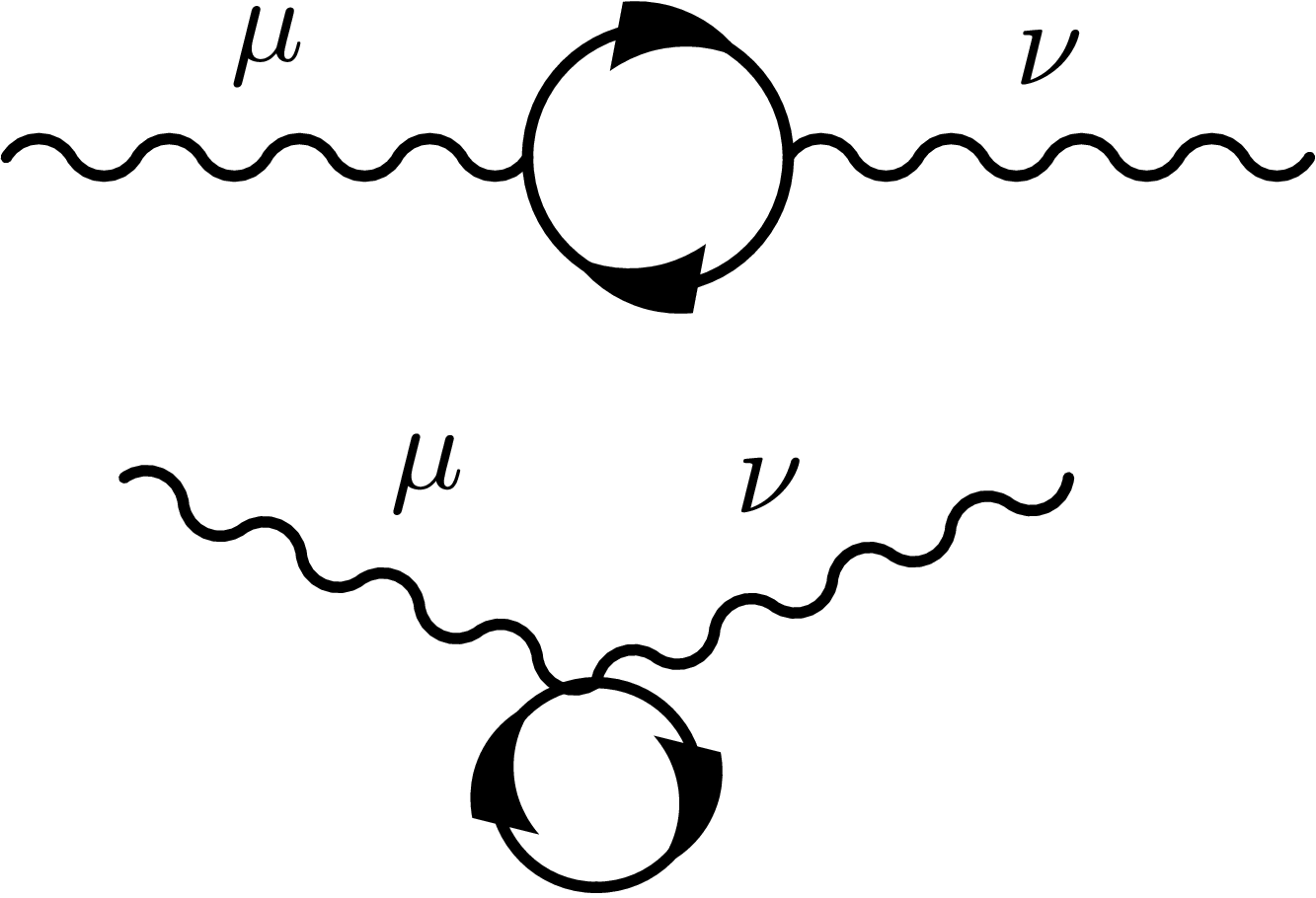}
  \caption{\label{fig:bubble} One-loop graphs contributing to the gauge field correlator in the large $N$ approximation.}
\end{figure}
In the large $N$ limit, this is determined by a one loop calculation depicted in Fig. \ref{fig:bubble}, which gives
\begin{equation}
\label{eq:largeNcor}
\Delta(q^2) = \frac{2\pi}{N}\left[\xi(q^2)\ln\left(\frac{\xi(q^2)+1}{\xi(q^2)-1}\right)-2\right]^{-1}
\end{equation}
where $\xi(q^2)=\sqrt{1+\frac{4M^2}{q^2}}$. 
In this example, the pure contact term approximation to the
topological charge correlator is obtained by assuming that the mass gap $M_0=2M$ is large compared to 
the momenta relevant to topological charge structure, i.e. $4M^2>>q^2$. Expanding in powers of $q^2/M^2$ and neglecting terms
that vanish in the limit $q^2/M^2\rightarrow 0$, we get
\begin{equation}
\label{eq:largeNcor2}
\Delta(q^2) =  \frac{12\pi M^2}{Nq^2} + \frac{6\pi}{5N} + O\left(\frac{q^2}{M^2}\right)
\end{equation} 
Note the presence of a $q^2=0$ pole in the gauge field correlator. Since the Chern-Simons current is
just the dual of the gauge field, the same massless pole also appears in the Chern-Simons correlator, and it's residue
determines the topological susceptibility in the large $N$ limit to be 
\begin{equation}
\chi_t =\frac{3M^2}{\pi N}+ O\left(\frac{1}{N^2}\right)
\end{equation}
When formulated in terms of the Chern-Simons correlator, 2D $\CP$ and 4D QCD 
have a similar topological charge structure at low 
momentum. In both cases, the topological susceptibility is just the
residue of the massless pole in the Chern-Simons current correlator. The massless Chern-Simons 
pole embodies the ``secret long range
order'' associated with topology in gauge theory \cite{Luscher78}. The physical significance of this massless pole
is subtle. Since the CS correlator is not gauge invariant, a $q^2=0$ pole does not imply the existence of a massless
particle. On the other hand, since the residue of the pole is the gauge invariant topological susceptibility,
the pole term in the CS current correlator is in fact gauge invariant and cannot be transformed away.
It thus represents a real physical long range correlation which is built into the gauge field by virtue of its 
topological fluctuations. Nevertheless, in the 
$q^2<<4M^2$ approximation, the topological charge correlator  
\begin{equation}
\tilde{G}(q^2) = \frac{1}{4\pi^2}q^2\Delta(q^2)
\end{equation}
is a polynomial in $q^2$, so that in coordinate space it collapses to local contact terms.
Numerically on the lattice, this leads to a $G(x)$ that is extremely short-range.

In the remainder of this section, we will discuss the contribution of small instantons 
to the correlator in the approximation of keeping only the contact terms in $\hat{G}(t)$. In this approximation, 
the time-dependence of the correlator is simply
\begin{equation}
\label{eq:tcorr}
\hat{G}(t) = -c_1\delta^{''}(t) + c_2\delta(t)
\end{equation}
We note that the numerical calculation of the topological susceptibility from the integral 
of the lattice correlator\cite{Ahmad} involves a large cancellation between the positive part of the correlator
at $t=0$ and 1, and the negative short-range tail from $t=2$ to $t\approx $ 4 or 5. A similar cancellation 
has been observed in 4D QCD \cite{Horvath-contact}. For the $\CP$ models, we find that, as beta is increased,
the range of the negative tail remains approximately fixed {\it in lattice units} as discussed in Section II.
This leads us to conclude that the negative tail of the 
correlator may be treated as part of the contact term, arising from the $\delta^{''}$ term 
in (\ref{eq:tcorr}).

\begin{figure}
  \centering
  \includegraphics[width=0.65\textwidth, angle=-90]{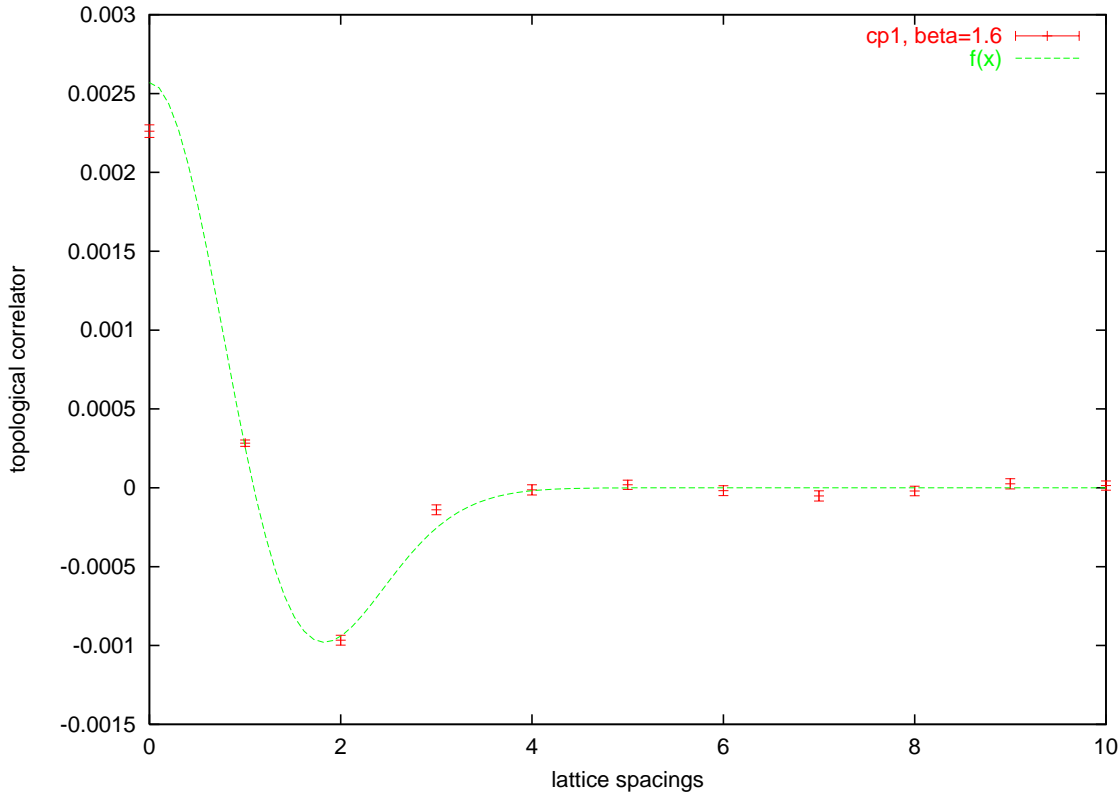}
  \caption{\label{fig:cp1_topcor_fit} Pure contact term fit to the two-point topological charge correlator in $CP^1$}
\end{figure} 

\begin{figure}
  \centering
  \includegraphics[width=0.65\textwidth, angle=0]{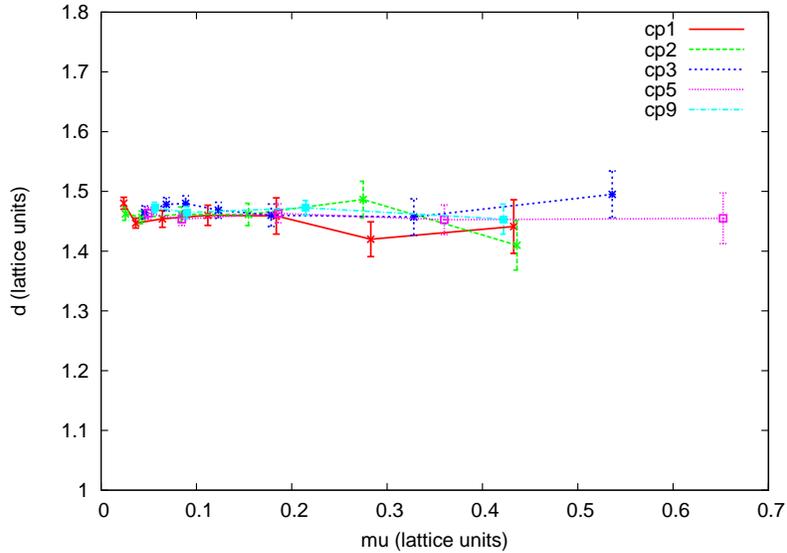}
  \caption{\label{fig:dparameter} The parameter $d$ defined in Eq. (\ref{eq:delta_function})}
\end{figure}

\begin{figure}
  \centering
  \includegraphics[width=0.65\textwidth, angle=0]{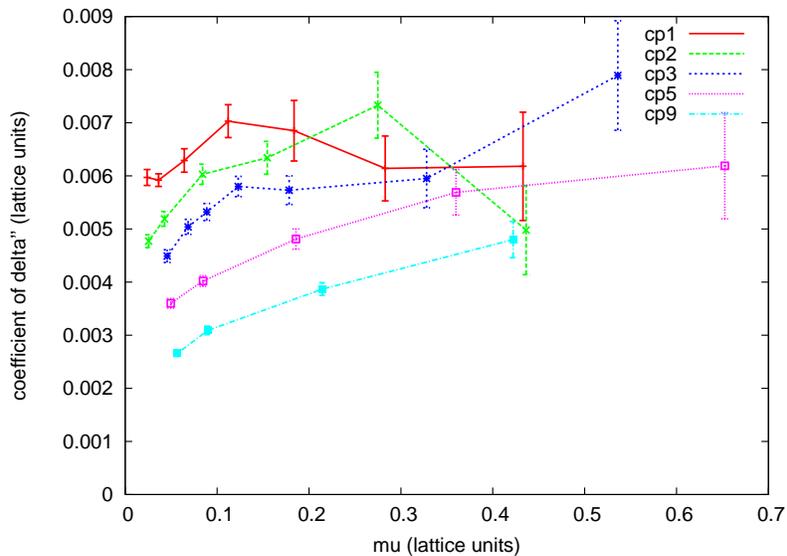}
  \caption{\label{fig:contact1} $\beta$ dependence of the OPE coefficient $c_1$ defined in Eq. (\ref{eq:tcorr})}
\end{figure}

Normal scaling behavior for the coefficients $c_1$ and $c_2$ would be
\begin{eqnarray}
c_1 \sim {\rm const.} \\
c_2 \sim \mu^2
\end{eqnarray}
In order to use Eq. (\ref{eq:tcorr}) as a fitting formula to extract the coefficients of the 
contact terms from the Monte Carlo data, we employ the following 
parametrization of a smeared delta function:
\begin{equation}
\label{eq:delta_function}
\delta(t) \rightarrow \frac{1}{d\sqrt{\pi}}e^{-t^2/d^2}
\end{equation}
The second derivative of this expression also provides a fitting function for the $\delta''(t)$ term in the 
correlator. We then fit the correlator to a sum of a $\delta(t)$ and a $\delta''(t)$ term. The parameters
which are allowed to vary in these fits are $c_1$, the coefficient of the $\delta''$ term, and $d$, the smearing
distance. The parameter $c_2$ is just the susceptibility $\chi_t$, which is computed separately and held fixed
in the fits. With these lattice approximations to $\delta(t)$ and $\delta''(t)$, the expression (\ref{eq:tcorr}) provides
a good fit to all of the correlators studied. Fig. \ref{fig:cp1_topcor_fit} shows a typical fit. This plot is for $CP^1$ at
$\beta=1.6$. The scaling behavior of the smearing parameter $d$ provides a significant test of the pure contact 
term assumption, Eq. (\ref{eq:tcorr}). In Fig. \ref{fig:dparameter} we plot the fit values for $d$ as a function of $\mu$
for various $\CP$ models. In all cases, $d$ approaches a finite value {\it in lattice units} of $d\approx 1.5$
lattice spacings. This is roughly the same as the measured thickness of domain walls reported in \cite{Ahmad}. Thus, the 
physical range of the correlator goes to zero in the continuum limit. This provides strong numerical support for
the assumption that the correlator is well-approximated by contact terms alone.
The scaling properties of $c_2=\chi_t$ for the various $\CP$ models have already been discussed. In particular, 
for $CP^1$ and $CP^2$, $c_2$ exhibits anomalous scaling properties due to small instantons. 
In Fig. \ref{fig:contact1} we have plotted the value of the $c_1$ coefficient as a function 
of $\mu$ for several $\CP$ models. From these results, we conclude that $c_1$ scales properly (i.e. approaches 
constant) as a function of $\mu$ for all the $\CP$ models. Most notably, unlike $c_2$, the coefficient 
$c_1$ does not appear to receive divergent contributions from small instantons, even for $CP^1$ and $CP^2$. 
The result that small instantons contribute only to the $c_2$ coefficient and not to the $c_1$ coefficient
is also expected on theoretical grounds, as we discuss in the next section. 
The fact that $c_1$ scales canonically (i.e. approaches a finite constant) in the continuum limit 
provides further support for the consistency of the pure contact term approximation
for the topological charge correlator. 

\section{Holographic and Conformal Properties of Topological Charge in $\CP$}
All of these results have a natural interpretation
in terms of a two-component picture for topological charge fluctuations in $\CP$ models, which we propose and discuss in
this Section.  To motivate this picture, we first observe that the required negativity of the $\langle q(x)q(0)\rangle$ 
correlator for nonzero separation has a fundamental dynamical effect which strongly constrains the types of
topological charge fluctuations which can contribute significantly to the vacuum path integral. The way that this 
constraint is realized in Monte Carlo configurations is by the predominance of topological charge distributions which have
a ``subdimensional'' character, i.e. ones in which coherent regions of topological charge are effectively either
zero-dimensional (small instantons) or one-dimensional (Wilson lines). A more unified perspective on both of these
types of topological charge excitations can be obtained by considering Witten's holographic formulation of 
theta dependence in gauge theory \cite{Witten98}. In this formulation, the theta term in 4D Yang-Mills 
arises from compactification of a 5D Chern-Simons term.
Similarly, we can interpret the theta term in $\CP$ as a compactified 3D Chern-Simons term,
\begin{equation}
{\cal L}_{CS} = i\epsilon^{abc}A_a \partial_b A_c
\end{equation}
Here, $a, b, c, \ldots$ run from 1 to 3.
Let us denote the original spacetime dimensions by 1 and 2, and the compactified dimension by 3.
Then in the limit of small radius of compactification, the Chern-Simons term reduces to a theta term,
\begin{equation}
{\cal L}_{CS} \rightarrow i\frac{\theta}{2\pi} \epsilon^{\mu\nu}\partial_{\mu}A_{\nu}=i\theta q(x)
\end{equation}
where $\mu,\nu = 1,2$, and
\begin{equation}
\label{eq:theta}
\theta = \oint A_3 dx_3
\end{equation}
From this 3-dimensional framework, a small instanton in $CP^1$ or $CP^2$ can be interpreted as a charged particle coupled
to the gauge field $A_3$ which has a world line wrapped around the compact direction in a closed loop, 
and is pointlike in the 1-2 plane. On the other hand,
we may integrate by parts and write,
\begin{equation}
{\cal L}_{CS} = -\frac{i}{2\pi} \epsilon_{\mu\nu}(\partial_{\mu}\theta)A_{\nu} \equiv J_{\nu}A_{\nu}
\end{equation}
where
\begin{equation}
\label{eq:dwcurrent}
J_{\nu} \equiv \frac{1}{2\pi}\epsilon_{\mu\nu}\partial_{\mu}\theta
\end{equation}
In this way of writing the CS term, the current $J_{\mu}$ couples to the gauge field in the 1-2 plane. 
In the limit of small compactification radius, the quantity $\theta$ defined by (\ref{eq:theta}) reduces
to the constant theta parameter of the 2D theory, but only mod $2\pi k$, where $k$ is an integer which labels a 
local k-vacuum state. These k-vacua are separated by domain walls, and the current $J_{\mu}$ is an ``edge current'' which
is nonvanishing along these domain walls.  

Thus, the 3D Chern-Simons/holographic view of theta dependence in $\CP$ models provides a more unified picture of the
two dominant types of topological charge excitations in $\CP$ models. 
In 3-dimensional spacetime, both small instantons and the domain walls associated with Wilson line excitations 
are 0-branes, i.e. particle-like
excitations with 1-dimensional world lines. They differ only in the orientation of their world line with respect to
the compactified dimension, with small instantons being wrapped around the compact direction localized in the 1-2 plane, 
and domain walls being world lines stretched out in some direction in the 1-2 plane. 
Some further insight into this structure,
which also exposes an apparent string/gauge connection,
can be gained by considering the 2D $\CP$ models in Lorentz gauge, $\partial_{\mu}A_{\mu}=0$. In this gauge, the 2D $A_{\mu}$
field may be written in terms of a single scalar field:
\begin{equation}
\label{eq:Lorentz}
A_{\mu} = \epsilon_{\nu\mu}\partial_{\nu}\Phi 
\end{equation}
The Chern-Simons current is the gradient of $\Phi$, and the topological charge is its Laplacian,
\begin{equation}
\label{eq:laplacian}
\nabla^2\Phi(x) = 2\pi q(x)
\end{equation}
The Monte Carlo results discussed here and in Ref. (\cite{Ahmad}) suggest the following idealization: Let us 
suppose that 
regions of nonzero topological charge in the vacuum are confined to zero- and one-dimensional submanifolds 
of two-dimensional spacetime. In this idealized view, spacetime consists mostly of voids in which $q(x)=0$,
with topological charge being confined to either isolated points (small instantons) or to boundaries between voids
(domain walls). 
In the voids $q(x)=0$, and (\ref{eq:laplacian}) implies that $\Phi(x_1,x_2)$, within a localized region of
spacetime and for a given gauge configuration, is a harmonic function, 
which can be written in terms of holomorphic and anti-holomorphic conformal fields, 
\begin{equation}
\label{eq:holo}
\Phi(x_1,x_2) = \phi(z)+\phi(\bar{z})
\end{equation}
where $z=x_1+ix_2$. In this framework, conformal symmetry is broken by the presence of instantons and domain walls
in the vacuum, represented respectively by poles and cuts in the analytic structure of $\phi(z)$ configurations
contributing to the path integral. The local conformal properties of the gauge field implied by this picture 
provide some insight into the collapsing of the OPE for the topological charge correlator discussed in 
Section IV. To illustrate this point,
consider the leading OPE term for the CS correlator (\ref{eq:OPE_CS}), which is conformally invariant. This vacuum insertion term
should be equal to the exact correlator in the absence of conformal symmetry breaking. The second term in 
(\ref{eq:OPE_CS}) is proportional to the (dimensionful) topological susceptibility, and thus arises from the 
breaking of scale invariance due to the presence of poles and cuts in the analytic structure of $\phi(z)$. 
In terms of the Lorentz gauge scalar field $\Phi$, 
the Chern-Simons current is $J_{\mu}^{CS}=\partial_{\mu}\Phi$, and the OPE coefficient $C_{1\mu\nu}(x)$ is
completely determined up to an overall constant by conformal invariance of the holomorphic and anti-holomorphic
terms in the correlator,
\begin{equation}
\label{eq:2point}
\langle\partial\phi(z_1)\partial\phi(z_2)\rangle = \frac{\rm const.}{(z_1-z_2)^2}
\end{equation}
and a similar expression for the anti-holomorphic part. The cross-correlators between a holomorphic and an 
anti-holomorphic field vanish. Using (\ref{eq:Lorentz}) and (\ref{eq:holo}),  
(\ref{eq:2point}) and the analogous correlator for $\partial_{\bar{z}}\phi(\bar{z})$ completely 
determine the form of the leading OPE coefficient
for the Chern-Simons correlator up to an overall constant,
\begin{equation}
C_{1\mu\nu}(x) = {\rm const.}\times\frac{x_{\mu}x_{\nu}-\frac{1}{2}x^2 \delta_{\mu\nu}}{(x^2)^2}
\end{equation}
It can now be seen that this OPE coefficient collapses upon differentiation, i.e.
\begin{equation}
\partial_{\mu}C_{1\mu\nu}(x) = {\rm const.}\times \partial_{\nu}\delta^2(x)
\end{equation}
This vacuum insertion term produces the $\delta^{''}(t)$ term in the $G(t)$ correlator but does not contribute
to $\chi_t$.

In the topological charge correlator, the $F^2$ insertion term (second term in the OPE (\ref{eq:OPE}) or (\ref{eq:OPE_CS})) breaks  
conformal invariance and provides nonvanishing $\chi_t$. Again, dimensional considerations and Lorentz invariance
allow us to write
\begin{equation}
\label{eq:C2_CFT}
C_{2\mu\nu} = {\rm const.}\times\left(\frac{1}{2}\delta_{\mu\nu}\ln x^2 + \frac{x_{\mu}x_{\nu}}{x^2}\right)
\end{equation}
(Here we have dropped an infrared divergent constant
$\propto \delta_{\mu\nu}\ln\lambda^2$, where $\lambda$ is an infrared cutoff mass.) The relative coefficient of the two
terms in (\ref{eq:C2_CFT}) is determined by the Lorentz gauge condition on the dual $A_{\mu}$ correlator.
The long range $\ln x^2$ behavior of the
Chern-Simons correlator coming from (\ref{eq:C2_CFT}) corresponds to the massless $1/q^2$ pole in the momentum space
correlator (\ref{eq:largeNcor}) or (\ref{eq:largeNcor2}). Thus, the large $x$ behavior of (\ref{eq:C2_CFT}) exhibits
the hidden long range order of the theory. Again, the corresponding
term in the topological charge correlator collapses to a contact term 
\begin{equation}
\label{eq:OPE2}
C_2(x) = -\partial_{\mu}\partial_{\nu}C_{2\mu\nu}(x) = {\rm const.}\times \delta^2(x)
\end{equation} 
It is this term in the correlator which gives nonzero topological susceptibility.

In Section IV it was shown numerically that small instantons 
only contribute to the $C_2$ coefficient and not to the $C_1$ coefficient in the OPE for the $q(x)$ correlator.
We can now present a theoretical argument for this by showing that the contribution of small instantons to the Chern-Simons
correlator is of precisely the form (\ref{eq:C2_CFT}). The distribution of Chern-Simons current around a small instanton
at spacetime point $y$ is given by 
\begin{equation}
J_{\mu}^{CS}(x) = \frac{1}{2\pi}\frac{(x-y)_{\mu}}{(x-y)^2}
\end{equation}
Thus, the contribution of an instanton to the correlator $\langle J_{\mu}^{CS}(x/2)J_{\nu}^{CS}(-x/2)\rangle$
is proportional to
\begin{equation}
\int \frac{d^2y}{(2\pi)^2}\frac{(y-\frac{1}{2}x)_{\mu}(y+\frac{1}{2}x)_{\nu}}{(y-\frac{1}{2}x)^2(y+\frac{1}{2}x)^2}
\equiv I_1+I_2
\end{equation}
where
\begin{equation}
I_1 =\int\frac{d^2y}{(2\pi)^2}\frac{y_{\mu}y_{\nu}}{(y-x/2)^2(y+x/2)^2}
\end{equation}
and
\begin{equation}
I_2 = -\frac{1}{4}x_{\mu}x_{\nu}\int\frac{d^2y}{(2\pi)^2}\frac{1}{(y-x/2)^2(y+x/2)^2}
\end{equation}
These integrals are of the same form as the one-loop Feynman integrals for the large-$N$ correlator (although here 
the integral is over instanton moduli space rather than momentum space). Again, as in (\ref{eq:C2_CFT}), we drop
infrared divergent constants which do not contribute to the topological charge correlator. This gives
\begin{equation}
I_1 = -\frac{1}{8\pi}\left(\delta_{\mu\nu}-\frac{x_{\mu}x_{\nu}}{x^2}\right)(\ln x^2 - 2)
\end{equation}
and 
\begin{equation}
I_2 = -\frac{1}{8\pi}\frac{x_{\mu}x_{\nu}}{x^2}\ln x^2
\end{equation}
Combining these (and dropping a constant term $\propto \delta_{\mu\nu})$ we get 
\begin{equation}
I_1+I_2 = -\frac{1}{4\pi}\left(\frac{1}{2}\delta_{\mu\nu}\ln x^2 + \frac{x_{\mu}x_{\nu}}{x^2}\right)
\end{equation}
Thus a small instanton gives a contribution to the correlator of precisely the same form as the $C_{2\mu\nu}$ coefficient, (\ref{eq:C2_CFT}).
We conclude that, within the dilute gas approximation, small instantons contribute only to the $c_2$ coefficient, in 
agreement with the numerical results of Section IV. 

\section{Conclusions}

Monte Carlo calculations using the overlap Dirac operator to study topological charge structure in $\CP$ models 
have begun to clarify some longstanding issues in these models. The $\CP$ models offer an ideal laboratory
for studying topological charge in asymptotically free gauge theory, 
and precise and detailed parallels can be drawn between these
models and QCD. Witten's original arguments that instantons in QCD will ``melt'' due to quantum fluctuations
\cite{Witten79} drew heavily on the $\CP$ analogy, in particular the absence of instanton effects in the large $N$
solution and nonzero topological susceptibility at order $1/N$. L\"{u}scher's arguments that topological charge in QCD could
be associated with ``Wilson bag'' surfaces were also supported by the $\CP$ analogy, where the corresponding excitation
is just the familiar Wilson line. Perhaps the most important simplification in the $\CP$ models, relative to
QCD, is the fact that in one spatial dimension, a domain wall is a point-like excitation. This allows the success of
standard field theory methods such as the large N expansion. The world lines of constituent $z$-particles which emerge in
the large $N$ limit can be identified with domain walls between k-vacua, where the value of $\theta/2\pi$, interpreted
as an electric field, changes by one unit of flux.  

For the $\CP$ models, a value of $N\approx 4$ marks the instanton melting point, i.e. the 
transition from dominance of $q(x)$ by 0-dimensional instantons
to dominance by one-dimensional line-like excitations. From a holographic 3D Chern-Simons viewpoint, the melting
of an instanton is actually the unwinding of a Wilson line excitation which was wrapped around the compact dimension.
Thus the line-like excitations which are found to dominate the larger N models might be viewed as not only a replacement
for the instantons, but in some sense the remnants of melted (unwound) instantons. 

\begin{figure}
  \centering
  \subfigure[$CP^1$]{
    \label{fig:cp1_chi_t_pm}
    \includegraphics[width=0.50\textwidth, angle=-90]{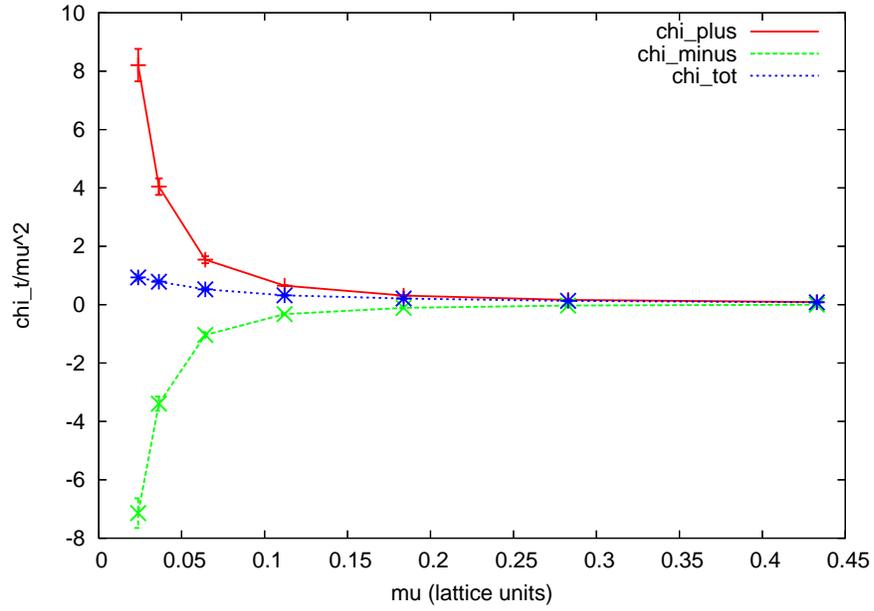}}
  \vspace{.1in}
  \subfigure[$CP^3$]{
    \label{fig:cp3_chi_t_pm}
    \includegraphics[width=0.50\textwidth, angle=-90]{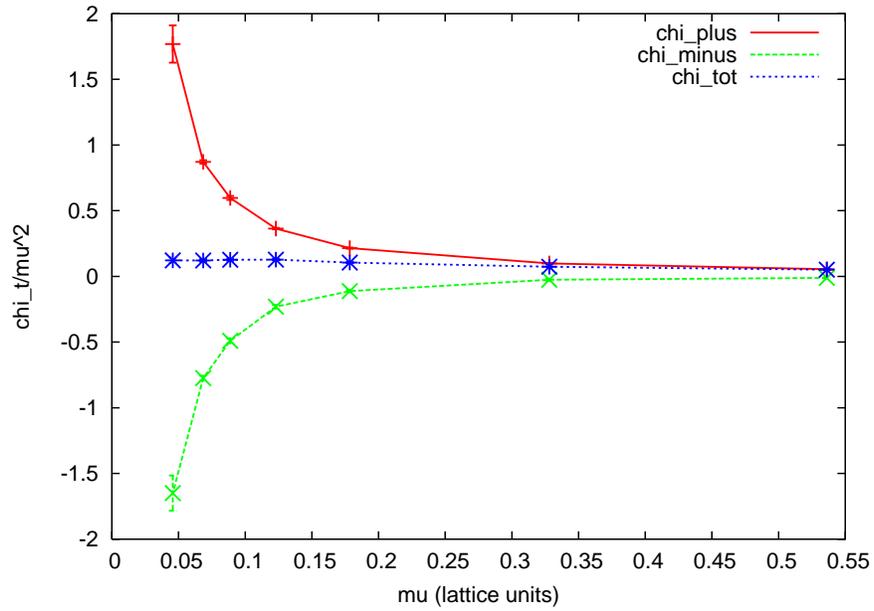}}
\caption{ Scaling behavior of the positive and negative contributions 
to the topological susceptibility in $CP^1$ and $CP^3$}
\end{figure}
 
The full dynamical implications of small instantons in the $CP^1$ and $CP^2$ models remain to be explored. One immediate
question is whether these models can have infinite topological susceptibility in the continuum
limit and still be consistent field theories. It is hard to imagine a more well-established quantum field theory than
the two-dimensional $O(3)$ sigma model, so we would be reluctant to conclude that this model cannot be 
defined as the continuum limit of a lattice model. What seems
more sensible is to regard the topological susceptibility as somewhat analogous to a mass parameter in a field theory
with two or more independent mass scales. At least in more benign cases, sending one mass to infinity removes a sector of
the theory but leaves the remainder of the theory well-defined. Such an interpretation of infinite topological
susceptibility is also suggested by a Witten-Veneziano type argument relating the magnitude of $\chi_t$ to the
splitting between the singlet and non-singlet mesons. By this argument, $z\bar{z}$ annihilation diagrams, 
allowed only in the singlet channel, will increase the mass$^2$ of the flavor singlet meson by an amount 
proportional to $\chi_t$. The $O(3)$ sigma model is an exactly solvable model, and it's particle spectrum is known
to consist of only an isotriplet of mesons, with no isosinglet. From the quark model point of view, this suggests
that the mass of the singlet has been shifted upward, if not by an infinite amount, 
then at least enough to make it unstable. By all indications,
the mass shift of the flavor singlet meson is the only singular physical effect of the divergent
$\chi_t$, and the resulting theory is in fact well-defined. Direct spectroscopic studies of the singlet channel
for various values of $N$ should help to clarify this issue. The lowest dimension operator that couples to the
singlet pseudoscalar channel is the topological charge $q(x)$. As we have seen, the two-point correlator for
$q(x)$ is dominated by contact terms and has little if any contribution from propagating intermediate states.
However, the coupling of $q(x)$ to singlet $z\bar{z}$ states may be suppressed, so it is difficult
to even put a lower bound on the singlet mass gap from the $q(x)$ correlator data alone. The study of other 
operators which couple to the pseudoscalar singlet channel (e.g. smeared $\bar{z}z$ operators) is essential
for a complete resolution of this issue.

The fact that the usual scaling $\chi_t\sim \mu^2$ is violated by the contribution of small instantons is not 
surprising, considering that in the limit of zero lattice spacing, a small instanton consists of a unit of 
topological charge within a vanishingly small radius, so that the topological charge density is becoming singular
in the continuum limit. The topological charge is much more spread out in the case of domain wall excitations
and should have a better chance of scaling properly. However, in both $\CP$ \cite{Ahmad} and in 
QCD \cite{Horvath:2003yj}, the domain walls are found to be of order lattice spacing in thickness, so that, while
they are less singular distributions than the small instantons, they are still not smooth on the lattice spacing
scale and would seem to have the potential to cause some problems with the scaling of topological susceptibility.
The naive argument that scaling occurs because everything becomes smooth and slowly varying at distances comparable
to the lattice spacing is clearly not applicable here. 
The numerical result that, for $CP^3$ and higher, topological susceptibility 
scales like $\mu^2$ and the physical $\chi_t$ approaches a finite value in
the continuum limit is thus not immediatedly obvious and probably deserves a deeper explanation. We note that
in calculating $\chi_t$ the integral under the positive peak from $0\leq x < 2$ and the integral of the negative
tail for $x\geq 2$ cancel with increasing accuracy as beta becomes large, as shown in Fig. \ref{fig:cp1_chi_t_pm}
and \ref{fig:cp3_chi_t_pm}.
This figure illustrates that, for the larger $N$ models, the positive and negative contributions to the correlator integral 
are separately diverging in the continuum limit,
but their sum scales properly to a finite topological susceptibility. For the $N>3$ models, this cancellation reflects the
dominance of the $\delta''(t)$ term, whose coefficient scales to a constant, compared to the coefficient of the $\delta(t)$
term, which scales $\propto \mu^2$.

One of our main motivations for undertaking a study of topological charge structure in the $\CP$ models was 
to try to explore the theoretical significance of the recently discovered membrane-like coherent topological
charge structure in QCD Monte Carlo configurations \cite{Horvath:2003yj}. As pointed out by L\"{u}scher \cite{Luscher82},
one can at least naively estimate the instanton melting point in 4-dimensional $SU(N)$ gauge theory. In Section
III we showed that with our lattice formulation of the model, the action of a small instanton was approximately
$\epsilon = \frac{N}{2}\times 6.8\ldots$
The critical value $\epsilon=4\pi$ gives an instanton melting point of $N_{crit}=3.7$. It may be possible to 
increase the small instanton action and thus lower $N_{crit}$ by improvement of the lattice $\CP$ action, 
but the lattice $\epsilon$ is bounded above by the continuum instanton action $\epsilon < 2\pi N$, so that
any lattice action for the $\CP$ model will give 
\begin{equation}
N_{crit} >2
\end{equation}
(Thus, in particular, no amount of improvement will completely eliminate small instantons from the $CP^1$ model.).
The lowest possible value of $N_{crit}=2$ can be estimated within the renormalized continuum theory 
alone, without reference to a lattice formulation. It
corresponds to the ``tipping point'' in the integral
over instanton radius in a semiclassical instanton calculation. This is where the integral changes from being
divergent at the small instanton end to being divergent at the large instanton end. (For example, the
integral over instanton radius in the $CP^1$ model goes like $\int d\rho/\rho$.) The analogous integral in
4-dimensional $SU(N)$ gauge theory (using the one-loop beta function) behaves like
\begin{equation}
\int \frac{d\rho}{\rho^5}\rho^{11N/3}
\end{equation}
which has it's tipping point at
\begin{equation}
\label{eq:Ncrit}
N_{crit} = \frac{12}{11}
\end{equation}
By this estimate, the value $N=3$ for real QCD is comfortably on the large-$N$ side of the critical value, but
(\ref{eq:Ncrit}) is only a lower bound on the value of $N_{crit}$ for any particular lattice action, so it
is not clear how seriously to take this estimate. The Monte Carlo evidence \cite{Horvath:2003yj} that membranes,
not instantons, dominate the topological charge distribution in QCD adds support for the notion that
$N=3$ QCD is in the large-$N$ regime, above the instanton melting point. As is well-known, the integral
over instanton radius in QCD is strongly weighted toward large instantons and must be cut off 
in an ad hoc way at about $0.3 fm$.
From the present point of view this is an indication that instantons are unstable toward melting or unwinding.
As L\"{u}scher originally argued 
\cite{Luscher78}, by periodicity in $\theta$, the vacuum inside a Wilson bag of charge $k=1$ has the same energy as 
the $\theta=0$ vacuum indicating that the
confining force between bag walls is completely screened by the appearance of an anti-bag. The analogous
process in the $\CP$ model is just ordinary string breaking and the vanishing of the area term for
integer-charged Wilson loops due to charge screening. With the confining force screened, 
the bag-antibag dipole layer is free
to expand and produce extended membranes with layers of topological charge of opposite sign juxtaposed, 
just the type of structure that is seen in the Monte Carlo configurations.


\end {document}